\documentclass[preprint]{vldb}
\usepackage{graphicx}
\usepackage{balance}  
\usepackage{algorithm}
\usepackage{bbm}
\usepackage{xspace}
\usepackage{xcolor}
\usepackage{enumitem}
\usepackage{booktabs}
\usepackage{listings}
\usepackage{hyperref}
\usepackage{lipsum}
\usepackage{setspace}
\usepackage[]{algpseudocode}
\usepackage[short]{optidef}

\algnewcommand\algorithmicforeach{\textbf{for each}}
\algdef{S}[FOR]{ForEach}[1]{\algorithmicforeach\ #1\ \algorithmicdo}

\newcommand{\cut}[1]{}
\newcommand{\sysname}{\textsc{PS$^3$}\xspace}

\newcommand{\minihead}[1]{{\vspace{.45em}\noindent\textbf{#1.} }}

\begin{document}


\title{Approximate Partition Selection for Big-Data Workloads using Summary Statistics}



%
%
%
%


\author{
Kexin Rong$^{\dag*}$, Yao Lu$^\dag$, Peter Bailis\footnotemark[1], Srikanth Kandula\footnotemark[2],  Philip Levis\footnotemark[1]\\\vspace{0.5em} 
\affaddr{Microsoft\footnotemark[2], Stanford\footnotemark[1]}
}

\maketitle

\begin{abstract}
Many big-data clusters store data in large partitions that support access at a coarse, partition-level granularity. As a result, approximate query processing via row-level sampling is inefficient, often requiring reads of many partitions. In this work, we seek to answer queries quickly and approximately by reading a subset of the data partitions and combining partial answers in a weighted manner \textcolor{black}{without modifying the data layout}. We illustrate how to efficiently perform this query processing using a set of pre-computed summary statistics, which inform the choice of partitions and weights. We develop novel means of using the statistics to assess the similarity and importance of partitions. Our experiments on several datasets and data layouts demonstrate that to achieve the same relative error compared to uniform partition sampling, our techniques offer from 2.7$\times$ to $70\times$ reduction in the number of partitions read, and the statistics stored per partition require fewer than 100KB.
\end{abstract}

\section{Introduction}
\label{sec:intro}
Approximate Query Processing (AQP) systems allow users to trade off between accuracy and query execution speed. In applications such as data exploration and visualization, this trade-off is not only acceptable but often desirable. Sampling is a common approximation technique, wherein the query is evaluated on a subset of the data, and much of the literature focuses on row-level samples~\cite{congressional,smallgroup,outlier}. 

\begin{sloppypar}
When data is stored in media that does not support random access~(e.g., flat files in data lakes and columnar stores~\cite{sqldw, hdfs}), constructing a row-level sample can be as expensive as scanning the entire dataset. For example, if data is split into partitions with $100$ rows, a $1$\% uniform row sample would in expectation require fetching $64$\% ($1-0.99^{100}$) of the partitions; a $10$\% uniform row sample would touch almost all partitions. As a result, recent work from a production AQP system shows that row-level sampling only offers significant speedups for complex queries where substantial query processing remains after the sampling~\cite{experiences}. 
\end{sloppypar}

\begin{figure}[t]
\includegraphics[width=\columnwidth]{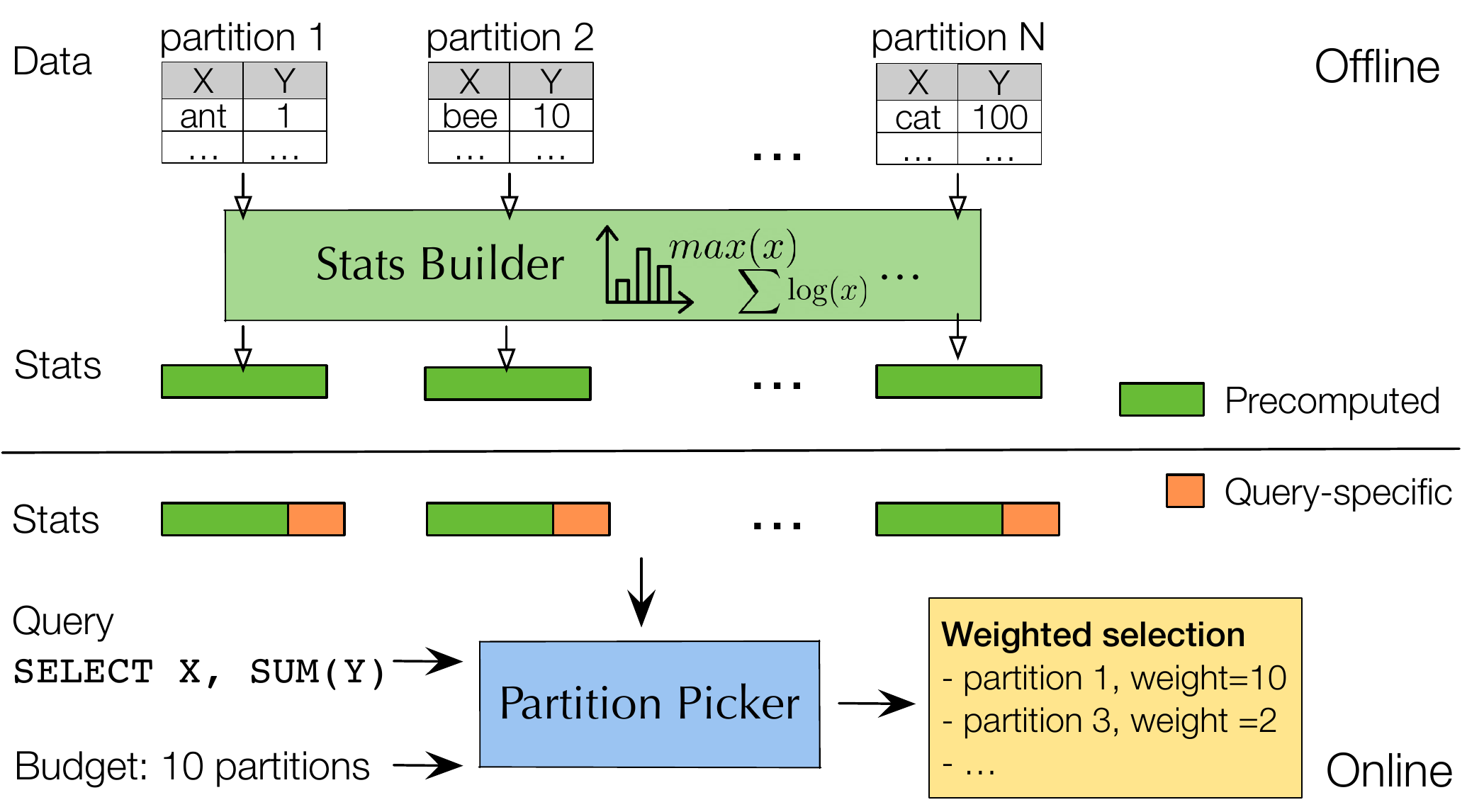}
\caption{Our system {\sysname} makes novel use of summary statistics to perform importance and similarity-aware sampling of partitions.\label{fig:arch}}
\end{figure}

In contrast to row-level sampling, the I/O cost of constructing a \emph{partition-level} sample is proportional to the sampling fraction\footnote{In the paper, we use ``partition'' to refer to the finest granularity at which the storage layer maintains statistics.}. In our example above, a 1\% partition-level sample would only read 1\% of the data. We are especially interested in big data clusters, where data is stored in chunks of tens to hundreds of megabytes, instead of disk blocks or pages which are typically a few kilobytes~\cite{gfs,hdfs}. Partition-level sampling is already used in production due to its appealing performance: commercial databases create statistics using partition samples~\cite{oracle, das2005block} and several Big Data stores allow sampling partitions of tables~\cite{hive, postgres, snowflake}. 

However, a key challenge remains in how to construct partition-level samples that can answer a given query accurately. Since all or none of the rows in a partition are included in the sample, the correlation between rows (e.g., due to layout) can lead to inaccurate answers. A uniformly random partition-level sample does not make a representative sample of the dataset unless the rows are randomly distributed among partitions~\cite{chaudhuri2004effective}, which happens rarely in practice~\cite{warehouse}. In addition, even a uniform random sample of rows can miss rare groups in the answer or miss the rows that contribute substantially to \texttt{SUM}-like aggregates. It is not known how to compute stratified~\cite{congressional} or measure-biased~\cite{sample+} samples over partitions, which helps with queries with group-by's and complex aggregates.

 \begin{sloppypar}
In this work, we introduce \sysname (Partition Selection with Summary Statistics), a system that supports AQP via weighted partition selection~(\autoref{fig:arch}). Our primary use case is in large-scale production query processing systems such as Spark \cite{sparksql}, F1~\cite{f1}, SCOPE~\cite{scope} where queries {\em only read} and datasets are {\em bulk appended}. Our goal is to minimize the approximation error given a sampling budget, or fraction of data that can be read. Motivated by observations from production clusters at Microsoft and in the literature that many datasets remain in the order that they were ingested~\cite{dip}, \sysname does not require any specific layout or re-partitioning of data. Instead of storing precomputed samples~\cite{blinkdb, smallgroup, chaudhuri2001robust}, which requires significant storage budgets to offer good approximations for a wide range of queries~\cite{bullet,quickr}, \sysname performs sampling exclusively during query optimization. Finally, similar to the query scope studied in prior work~\cite{blinkdb,dblearning,skipping}, \sysname supports single-table queries with \texttt{SUM, COUNT(*), AVG} aggregates, \texttt{GROUP BY} on columnsets with moderate distinctiveness, predicates that are conjunctions, disjunctions or negations over single-column clauses. 
\end{sloppypar}

To select partitions that are most relevant to a query, \sysname leverages the insight that partition-level summary statistics are relatively inexpensive to compute and store. The key question is which statistics to use. Systems such as Spark SQL and ZoneMaps already maintain statistics such as maximum and minimum values of a column to assist in query optimization~\cite{dip}. Following similar design considerations, we look for statistics with small space requirements that can be computed for each partition in one pass at ingest time. For functionality, we look for statistics that are discriminative enough to support decisions such as whether the partition contributes disproportionally large values of the aggregates. We propose such a set of statistics for partition sampling -- measures, heavy hitters, distinct values, and histograms -- which include but expand on conventional catalog-level statistics. The total storage overhead scales with the number of partitions instead of with the dataset size. We only maintain single-column statistics to keep the overhead low. The resulting storage overhead can be orders of magnitudes smaller than approaches using auxiliary indices to reduce the cost of random access~\cite{sample+}. While the set of statistics is by no means complete, we show that each type of statistics contributes to the sampling performance, and, in aggregate, delivers effective AQP results.

We illustrate three ways of using summary statistics to help with partition selection. First, if we knew which partitions contribute more to the query answer, we could sample these partitions more frequently. While it is challenging to manually design rules that relate summary statistics to partition contribution, a model can \emph{learn} how much summary statistics matter through examples. Inspired by prior work which uses learning techniques to improve the sampling efficiency for counting queries on row-level samples~\cite{lss}, we propose a learned importance-style sampling algorithm that works on aggregate queries with \texttt{GROUP BY} clauses and on partitions. The summary statistics serve as a natural feature representation for partitions, from which we can train models offline to learn a mapping from summary statistics to the relative importance of a partition. During query optimization, we use the trained models to classify partitions into several importance groups, and allocate the sampling budget across groups such that the more important groups get a greater proportion of the budget. The training overhead is a one time cost for each dataset and workload, and for high-value datasets in clusters that are frequently queried, this overhead is amortized over time. 

In addition, we leverage the redundancy and skewness of the partitions for further optimization. For two partitions that output similar answers to an input query, it suffices to only include one of them in the sample. While directly comparing the contents of the two partitions is expensive, we can use the query-specific summary statistics as a proxy for the similarity between partitions. We also observe that datasets commonly exhibit significant skew in practice. For example, in a prototypical production service request log dataset at Microsoft, the most popular application version out of the 167 distinct versions accounts for almost half of the dataset. Inspired by prior works in AQP that recognize the importance of outliers~\cite{smallgroup,outlier}, we use summary statistics (e.g., the occurrences of heavy hitters in a partition) to identify a small number of partitions that are likely to contain rare groups, and dedicate a portion of the sampling budget to evaluate these partitions exactly.

In summary, this paper makes the following contributions:
\begin{enumerate}[topsep=2pt,parsep=1pt]

\item We introduce \sysname, a system that makes novel uses of summary statistics to perform weighted partition selection for many popular queries. Given the query semantics, summary statistics and a sampling budget, the system intelligently combines a few sampling techniques to produce a set of partitions to sample and the weight of each partition.

\item We propose a set of lightweight sketches for data partitions that are not only practical to implement, but can also produce rich partition summary statistics. While the sketches are well known, this is the first time the statistics are used for weighted partition selection.

\item We evaluate on a number of real-world datasets with real and synthetic workloads. Our evaluation shows that each component of \sysname contributes meaningfully to the final accuracy and together, the system outperforms alternatives across datasets and layouts, delivering from 2.7$\times$ to $70\times$ reduction in data read given the same error compared to uniform partition sampling. 
\end{enumerate}

\section{System Overview}
In this section, we give an overview of \sysname, including its major design considerations, supported queries, inputs, and outputs, and the problem statement. 
 
\subsection{Design Considerations}
\label{sec:design}
We highlight a few design considerations in the system.  

\minihead{Layout Agnostic} A random data layout would make the partition selection problem trivial, but maintaining a random layout requires additional efforts and rarely happens in practice~\cite{warehouse}. In read-only or append-only data stores, it is also expensive to modify the data layout. As a result, we observe that in practice, many datasets simply remain in the order that they were ingested in the cluster. In addition, prior work~\cite{skipping} has shown that it is challenging and sometimes impossible to find a partitioning scheme that enables good data skipping for arbitrary input queries. Therefore, instead of requiring re-partitioning or random layout, \sysname explicitly chooses to keep data in situ and tries to make the best out of the given data layout. We show that \sysname can work across different data layouts in the evaluation (\S~\ref{sec:dl}). 

\minihead{Sampling on a single table} To perform joins effectively, prior work~\cite{quickr} has shown that sampling each input relation independently is not enough and that the joint distribution must be taken into account. Handling the correlations between join tables at the partition level is another research problem on its own, and is outside the scope of this paper. However, sampling on a single table can still offer non-trivial performance improvements even for queries that involve joining multiple tables. For example, in key--foreign key joins, fact tables are often much larger compared to dimension tables. Sampling the fact table, therefore, already gets us most of the gains.

\minihead{Generalization} Prior works make various trade-offs between the efficiency and the generality of the queries that they support, ranging from having access to the entire workload~\cite{skipping} to being workload agnostic~\cite{ola}. Our system falls in the middle of the spectrum, where we make assumptions about the structure and distribution of the query workload. Specifically, we assume that the set of columns used in \texttt{GROUP BY}s and the aggregate functions are known apriori, with the scope defined in \S~\ref{sec:support}; predicates can take any form that fits under the defined scope and we do not assume we have access to the exact set of predicates used. We assume that a workload consists of queries made of an arbitrary combination of aggregates, group bys and predicates from the scope of interest. \sysname is trained per data layout and workload, and generalizes to unseen queries sampled from the same distribution as the training workload. Overall, our system is best suited for commonly occurring queries and should be retrained in case of major changes in query workloads such as the introduction of unseen group by columns.

We do not consider generalization to unseen data layouts or datasets and we view broader generalization as an exciting area for future work (\S~\ref{sec:future}). Since most summary statistics are computed per column, different datasets might not share any common statistics. Even for the same dataset, the importance of summary statistics can vary across data layouts. For example, the mean of column $X$ can distinguish partitions in a layout where the dataset is sorted by $X$, but may provide no information in a random layout.

\subsection{Supported Queries}
\label{sec:support}
In this section, we define the scope of queries that \sysname supports. We support queries with an arbitrary combination of aggregates, predicates and group bys. Although we do not directly support nested queries, many queries can be flattened using intermediate views~\cite{views}. Our techniques can also be used directly on the inner queries. Overall, our query scope covers 11 out of 22 queries in the TPC-H workload (Appendix A.1, extended report~\cite{app}). 

\begin{itemize}[topsep=0.5pt,itemsep=0.5pt,parsep=0.5pt]
    \item  \texttt{Aggregates}. We support \texttt{SUM} and \texttt{COUNT(*)} (hence \texttt{AVG}) aggregates on columns as well as simple linear projections of columns in the select clause. The projections include simple arithmetic operations (\texttt{+, -}) on one or more columns in the table\footnote{We also support the multiply and divide operations in some cases using statistics computed over the logs of the columns.}. We also support a subset of aggregates with \texttt{CASE} conditions that can be rewritten as an aggregate over a predicate.
    \item \texttt{Predicates}. Predicates include conjunctions, disjunctions and negations over the clauses of the form $c$ op $v$, where $c$ denotes a column, op an operation and $v$ a value. We support equality and inequality comparisons on numerical and date columns, equality check with a value as well as  the \texttt{IN} operator for string and categorical columns as clauses. 
    \item \texttt{Groups}. We support \texttt{GROUP BY} clauses on one or more stored attributes\footnote{To support derived attributes, we make a new column from the derived attribute and store its summary statistics}. We do not support \texttt{GROUP BY} on columns with large cardinality since there is little gain from answering highly distinct queries over samples; one could either hardly perform any sampling without missing groups, or would only care about a limited number of groups with large aggregate values (e.g., \texttt{TOP} queries), which is out of the scope of this paper. 
  \item \texttt{Joins}. Queries containing key--foreign joins can be supported as queries over the corresponding denormalized table. For simplicity, our discussion in this paper is based on a denormalized table.
\end{itemize}

In the TPC-H workload, 16 out of the 22 queries can be rewritten on a denormalized table and 11 out of the 16 are supported by our query scope. For the 5 that are not supported, 4 involve group bys on high cardinality columns and 1 involves the \texttt{MAX} aggregate. A number of prior work have also studied similar query scopes~\cite{blinkdb, dblearning, skipping}.

\subsection{Inputs and Outputs}
 \begin{sloppypar}
\sysname consists of two main components: the statistics builder and the partition picker (\autoref{fig:arch}).  In this section, we give an overview of the inputs and outputs of each component during preprocessing and query time. 
 \end{sloppypar}

\subsubsection{Statistics Builder}
\minihead{Preparation} The statistics builder takes a data partition as input and outputs a number of lightweight sketches for each partition. The sketches are stored separately from the partitions. We describe the sketches used in detail, including the time and space complexity for constructing and storing the sketches in \S~\ref{sec:sketches}.

\minihead{Query Time} During query optimization, one can access the sketches {\em without} touching the raw data. Given an input query, the statistics builder combines pre-computed column statistics with query-specific statistics computed using the stored sketches and produces a set of summary statistics for each partition and for each column used in the query. 

\subsubsection{Partition Picker}

\minihead{Preparation} In the preparation phase, the picker takes a specification of workload in the form of a list of aggregate functions and columnsets that are used in the \texttt{GROUP BY}. \textcolor{black}{We can sample a query from the workload by combining randomly generated predicates and randomly selected aggregate functions and group by columnsets (0 or 1) from the specification. For each sampled query, we compute the summary statistics as well as the answer to the query on each partition as the training data, which the picker uses to learn the relevance of different summary statistics. The training is a one time cost and we train one model for each workload to be used for all test queries. We elaborate on the design of the picker in \S~\ref{sec:algo}}. 

\minihead{Query Time} The picker takes an input query, summary statistics and a sampling budget as inputs, and outputs a list of partitions to sample, as well as the weight of each partition in the sample. This has a net effect of replacing a table in the query execution plan with a set of weighted partition choices with a small overhead (\autoref{tab:latency}). Query execution should also be augmented to handle weights, similar to modifications suggested in prior work~\cite{quickr}.

\subsection{Problem Statement}
\label{sec:prob}
Let $N$ be the total number of partitions and $M$ be the dimension of the summary statistics. For an aggregation query $Q$, let $G$ be the set of groups in the answer to $Q$. For each group $g \in G$, denote the aggregate values for the group as ${\bf A_{g}} \in \mathbbm{R}^d$, where $d$ is the number of the aggregates. Denote the aggregates for group $g$ on partition $i$ as ${\bf A_{g, i}}$.

Given the input query $Q$, the summary statistics $F \in \mathbbm{R}^{N\times M}$ as well as sampling budget $n$ in the form of number of partitions to read, our system returns a set of weighted partition choices $S = \{(p_1, w_1), (p_2, w_2), ..., (p_n, w_n)\}$. The approximate answer ${\bf \tilde{A}_g}$ of group $g$ for $Q$ is computed by
${\bf \tilde{A}_g} = \sum_{j=1}^n  w_j {\bf A_{g,p_j}}, \forall g \in G$. 

Our goal is to produce the set of weighted partition choice $S$ such that ${\bf \tilde{A}_g}$ is a good approximation of the true answer ${\bf A_g}$ for all groups $g \in G$. To assess the approximation quality across groups and aggregates that are of different sizes and magnitudes, we measure absolute and relative error, as well as the percentage of groups that are missed in the estimate. 
\section{Partition Summary Statistics}
\label{sec:features}
The high-level insight of our approach is that we want to differentiate partitions based on their contribution to the query answer, and that the contribution can be estimated using a rich set of partition-level summary statistics. As a simple example, for \texttt{SUM}-type aggregates, partitions with a higher average value of the aggregate should be preferred, all else being equal. We are unaware of prior work that uses partition-level summary statistics for performing non-uniform partition selection. In this section, we describe the design and implementation of the summary statistics.

\begin{table}[t]
\begin{center}
\small
\caption{Per partition, the time and space overheads to construct and store sketches for partitions with $R_b$ rows. Small logarithmic factors are ignored. }
\begin{tabular}{ l l l }
\toprule
 \textbf{Sketch} & \textbf{Construction} & \textbf{Storage} \\ 
 \midrule
  Histograms & $O(R_b\log{R_b})$ & $O(\#buckets)$\\
 Measures & $O(R_b)$ & $O(1)$\\
 AKMV & $O(R_b)$ & $O(k)$\\
 Heavy Hitter & $O(R_b)$ & $O(\frac{1}{support})$ \\
 \bottomrule
\end{tabular}
\label{tab:sketch}
\end{center}
\vspace{-1.5em}
\end{table}

\subsection{Lightweight Sketches}
\label{sec:sketches}
Our primary use case, similar to columnar databases, is read-only or append-only stores. Summary statistics are constructed for each new data partition when the partition is sealed. The necessary data statistics should be simple, small in size and can be computed incrementally in one pass over data. The necessary statistics should also be discriminative enough to set partitions apart and rich enough to support sampling decisions such as estimating the number of rows that pass the predicate in a partition. We opt to use only single-column statistics to keep the memory overhead light, although more expensive statistics such as multi-column histograms can help estimate selectivity more accurately. The design considerations lead us to the following sketches:
\begin{sloppypar}
\begin{itemize}[itemsep=0.5pt,parsep=0.5pt]
    \item \textbf{Measures:} Minimum, maximum, as well as first and second moments are stored for each numeric column. For columns whose value is always positive, we also store measures on the log transformed column. 
    \item \textbf{Histogram:} We construct equal-depth histograms for each column. For string columns, the histogram is built over hashes of the strings. By default, each histogram has 10 buckets. 
    \item \textbf{AKMV:}  We use an AKMV (K-Minimum Values) sketch to estimate the number of distinct values~\cite{kmv}. The sketch keeps track of the $k$ minimum hashed values of a column and the number of times these values appeared in the partition. We use $k=128$ by default. 
    \item \textbf{Heavy Hitter:} We maintain a dictionary of heavy hitters and their frequencies for each column in the partition using lossy counting~\cite{hh}. By default, we only track heavy hitters that appear in at least 1\% of the rows, so the dictionary has at most 100 items.
\end{itemize}
\end{sloppypar}
\autoref{tab:sketch} summarizes the time complexity to construct the sketches and the space overhead to store them, ignoring small logarithmic factors. The sketches can be constructed in parallel for each partition. We do not claim that the above choices make a complete set of sketches that should be used for the purpose of partition selection. Our point is that these are a set of inexpensive sketches that can be easily deployed or might have already been maintained in big-data systems~\cite{dip}, and that they can be used in new ways to improve partition sampling. 

\begin{table}[t]
\begin{center}
\small
\caption{Summary statistics and the sketches used to compute them. Selectivity is computed per query and all other statistics is computed per column. }
\begin{tabular}{ l l }
\toprule
 \textbf{Summary Statistics} & \textbf{Sketch} \\ 
 \midrule
 $\overline{x}$, $min(x)$, $max(x)$, $\overline{x^2}$, $std(x)$ & Measures  \\  
 $\overline{\log(x)}$,  $\overline{\log(x)^2}$, $min(\log(x))$, $max(\log(x))$ & Measures \\
 number of distinct values & AKMV  \\
 avg/max/min/sum freq. of distinct values & AKMV  \\
 \# hh, avg/max freq. of hh & Heavy Hitter  \\
occurrence bitmap of heavy hitters  & Heavy Hitter \\
 selectivity & Histogram\\
 \bottomrule
\end{tabular}
\label{tab:feat}
\end{center}
\vspace{-1.5em}
\end{table}

\subsection{Summary Statistics as Features} 
\label{sec:stats}
Given the set of sketches, we compute summary statistics for each partition, which can be used as {\em feature vectors} to discriminate partitions based on their contribution to the answer of a given query. The features consist of two parts: pre-computed per column features and query-specific selectivity estimates (\autoref{tab:feat}). We apply a query-dependent mask on the pre-computed column features: features associated with columns that are unused in the query are set to zero. In addition, for categorical columns where the measure based sketches do not apply, we set the corresponding features to zero. The schema of the feature vector is determined entirely by the schema of the table, so queries on the same dataset share the same feature vector schema.

Overall, there are four types of features based on the underlying sketches that generate them: measures, heavy hitters, distinct values and selectivity. Each type of feature captures different information about the partitions and the queries. Measures help identify partitions with disproportionally large values of the aggregates; heavy hitters and distinct values help discriminate partitions from each other and selectivity helps assess the impact of the predicates. We found that all types of features are useful in \sysname but the relative importance of each varies across datasets (\S~\ref{sec:featimp}). 

Extracting features from sketches is, in general, straightforward; we discuss two interesting cases below.

\minihead{Occurrence Bitmap} We found that it is not only helpful to know the number of heavy hitters, but also \emph{which} heavy hitters are present in the partition. To do so, we collect a set of $k$ global heavy hitters for a column by combining the heavy hitters from each partition. For each partition, we compute a bitmap of size $k$, each bit representing whether the corresponding global heavy hitter is also a heavy hitter in the current partition. The feature is only computed for grouping columns and we cap $k$ at 25 for each column.  

\minihead{Selectivity Estimates} The selectivity estimate is a real number between 0 and 1, designed to reflect the fraction of rows in the partition that satisfies the query predicate. The estimate supports predicates defined in our query scope (\S~\ref{sec:support}) and is derived using histograms over individual columns. Predicate clauses that use the same column (e.g., $X<1$ or $X>10$) are evaluated jointly. As a special case, if a string column has a small number of distinct values, all distinct values and their frequencies are stored exactly; this can support regex-style textual filters on the string column (e.g., \texttt{'\%promo\%'}). We use the following four features to represent the selectivity of predicates which can be a conjunction or disjunction of individual clauses:
\begin{sloppypar}
\begin{enumerate}[itemsep=0.5pt,parsep=0.5pt]
    \item \texttt{selectivity\_upper}: For \texttt{AND}s, the selectivity is at most the min of the selectivity of individual clauses; for \texttt{OR}s, the selectivity is at most 1 and at most the sum of the selectivity of individual clauses. 
    \item \texttt{selectivity\_indep}: This feature computes the selectivity assuming independence between predicate clauses. For \texttt{AND}s, the feature is the product of the selectivity for each individual clause; for \texttt{OR}s, the feature is the min of the selectivity of individual clauses. 
      \item \texttt{selectivity\_min}, \texttt{selectivity\_max}: We store the min and max of the selectivity of individual clauses.
\end{enumerate}
\end{sloppypar}

If the upper bound of the selectivity is zero, the partition contains no rows that pass the predicate; if the upper bound is nonzero however, the partition can have zero or more rows that pass the predicate. In other words, as a classifier for identifying partitions that satisfy the predicate, \texttt{selectivity\_upper}$>0$ has perfect recall and uncertain precision. For simple predicates such as $X > 1$, the precision is 100\%; for complicated predicates involving conjunctions and disjunctions over many clauses and columns (e.g., TPC-H Q19), the precision can be as low as 10\%.

\section{Partition Picking}
\label{sec:algo}
In this section, we describe \sysname's partition picker component and how it makes novel use of the summary statistics discussed above to realize weighted partition selection. 

\subsection{Picker Overview}
To start, we give an overview of how our partition picker works. Recall that the picker takes a query, the summary statistics and a sampling budget as inputs, and outputs a list of partitions to evaluate the query on and the weight of each partition. Partial answers from the selected partitions are combined in a weighted manner, as described in \S~\ref{sec:prob}.

Algorithm~\ref{alg:test} describes the entire procedure. We first identify outlier partitions with rare groups using the procedure described in \S~\ref{sec:outlier}. Each outlier partition has a weight of 1. We then use the trained models to classify the remaining partitions into groups of different importance, using the algorithm described in \S~\ref{sec:clf}. We allocate the remaining sampling budget across groups such that the sampling rate decreases by a factor of $\alpha$ from the $i^{th}$ important to the $(i+1)^{th}$ important group. Finally, given a sample size and a set of partitions in each importance group, we select samples via clustering using the procedure described in \S~\ref{sec:cluster}. An exemplar partition is selected from each cluster, and the weight of the exemplar equals the size of the cluster. We explain each component in detail in the following sections. 

\begin{algorithm}[t]
\caption{Partition Picker}
\label{alg:test}
\begin{algorithmic}[1]
\Require{partition features $F$, sampling budget $n$, group-by columns {\sf gb\_col}, models {\sf regrs}, decay rate $\alpha$}
\Ensure{{\sf selection}: [$(p_1, w_1), (p_2, w_2), ..., (p_n, w_n)$]}
\State {\sf outliers, inliers} $\gets $ \textproc{Outlier}($F$, {\sf gb\_col})
\State $n_o \gets$ {\sf outliers}.size()
\State {\sf selection}.add({\sf outliers}, $[1] * n_o$)
\State {\sf groups} $\gets$ \textproc{ImportanceGroup}($F$, {\sf inliers, regrs})
\State $n_c \gets $  \textproc{AllocateSamples}({\sf groups}, $n - n_o$, $\alpha$)
\For {$i \gets 1, ..., ${\sf groups}.size()}
\State {\sf selection}.add(\textproc{Clustering}($F$[{\sf groups[i]}], $n_c[i]$)) 
\EndFor
\end{algorithmic}
\end{algorithm}

\subsection{Sample via Clustering}
\label{sec:cluster}
We start by describing the sample selection procedure (line 7 in Algorithm~\ref{alg:test}), designed to leverage the redundancy between partitions. We use feature vectors to compute a similarity score between partitions, which consequently enables us to choose dissimilar partitions as representatives of the dataset. In fact, identical partitions will have identical summary statistics, but the converse does not hold; having summary statistics on multiple columns as well as multiple statistics for each column makes it less likely that dissimilar partitions have identical summary statistics. 

We propose to use {\em clustering} as a sampling strategy: given a sampling budget of $n$ partitions, we perform clustering using feature vectors with a target number of $n$ clusters; an exemplar partition is chosen per cluster, with an assigned weight equals the number of partitions in the cluster. Denote the answer to the query on cluster $i$'s exemplar partition as $A_i$ and the size of cluster $i$ as $s_i$. The estimate of the query answer is given by $\Tilde{A} = \sum_{i=1}^{n}  s_iA_i $. 

\begin{sloppypar}Concretely, we measure partition similarity using Euclidean distances of the feature vectors. We zero out features for unused columns in the query so they have no impact on the result; we also perform normalization such that the distance is not dominated by any single feature (Appendix B). Regarding the choice of the clustering algorithm, we experimented with KMeans and Agglomerative Clustering and found that they perform similarly. Finally, the cluster exemplar is selected by picking the partition whose feature vector has the smallest distance to the median feature vector of partitions in the clusters. \end{sloppypar}

Our proposed scheme leads to a biased estimator that can be challenging to analyze. Specifically, given the median feature vector of a cluster, our estimator deterministically picks the partition that is closest to the median vector as the cluster exemplar. However, one could make a simple modification to unbias the estimator by selecting a random partition in the cluster as the exemplar instead. We have included an empirical comparison of the accuracy of the two estimators as well as a variance analysis for the unbiased estimator in Appendix D. We have empirically found that the proposed scheme outperforms its unbiased counterpart when the sampling budget is limited.

Clustering effectively leverages the redundancy between partitions, especially in cases when partitions have near identical features. Although there is no guard against an adversary, in practice, having a large and diverse set of summary statistics makes it naturally difficult for dissimilar partitions to be in the same cluster. Clusters play a similar role as strata in stratified sampling. The goal of clustering is to make partitions in the same stratum homogeneous such that the overall sampling variance is reduced. Finally, clustering results vary from query to query: the same partition can be in different clusters for different queries due to the changes in selectivity features and the query-dependent column masks.

\minihead{Feature Selection} Clustering assumes that all features are equally relevant to partition similarity. To further improve the clustering performance, we perform feature selection via a ``leave-one-out" style test. For example, consider a table with columns $X, Y$ and features $min, max$. We compare the clustering performance on the training set using $\{min(X), max(X), min(Y), max(Y)\}$ as features to that from using only $\{max(X), max(Y)\}$ as features. If the latter gives a smaller error, we subsequently exclude the $min$ feature for all columns from clustering. We greedily remove features until converging to a local optimal, at which point excluding any remaining features would hurt clustering performance. In an outer loop, we repeat the above greedy procedure multiple times, each time starting with a random ordering of the features. Our experiments show that feature selection consistently improves clustering performance across datasets. We provide the pseudo code of the procedure in Appendix B.1. 

\minihead{Limitations}  We briefly discuss two failure cases for clustering in which \sysname can fall back to random sampling (details in Appendix B.1).  First, clustering takes advantage of the redundancy among partitions. In the extreme case when the query groups by the primary key, no two partitions contribute similarly to the query and any downsampling would result in missed groups. As discussed in \S~\ref{sec:support}, our focus is on queries where such redundancy exists. Second, queries with highly selective predicates might suffer from poor clustering performances. Since most features are computed on the entire partition, the features would no longer be representative of partition similarity if only a few rows satisfy the predicate in each partition. 

\subsection{Learned Importance-Style Sampling}
\label{sec:clf}
While clustering helps select partitions that are dissimilar, it makes no distinction between partitions that contribute more to the query and partitions that contribute less. Ideally, we would want to sample the more important partitions more frequently to reduce the variance of the estimate~\cite{is}.

\begin{figure}[t]
\centering
\includegraphics[width=0.9\columnwidth]{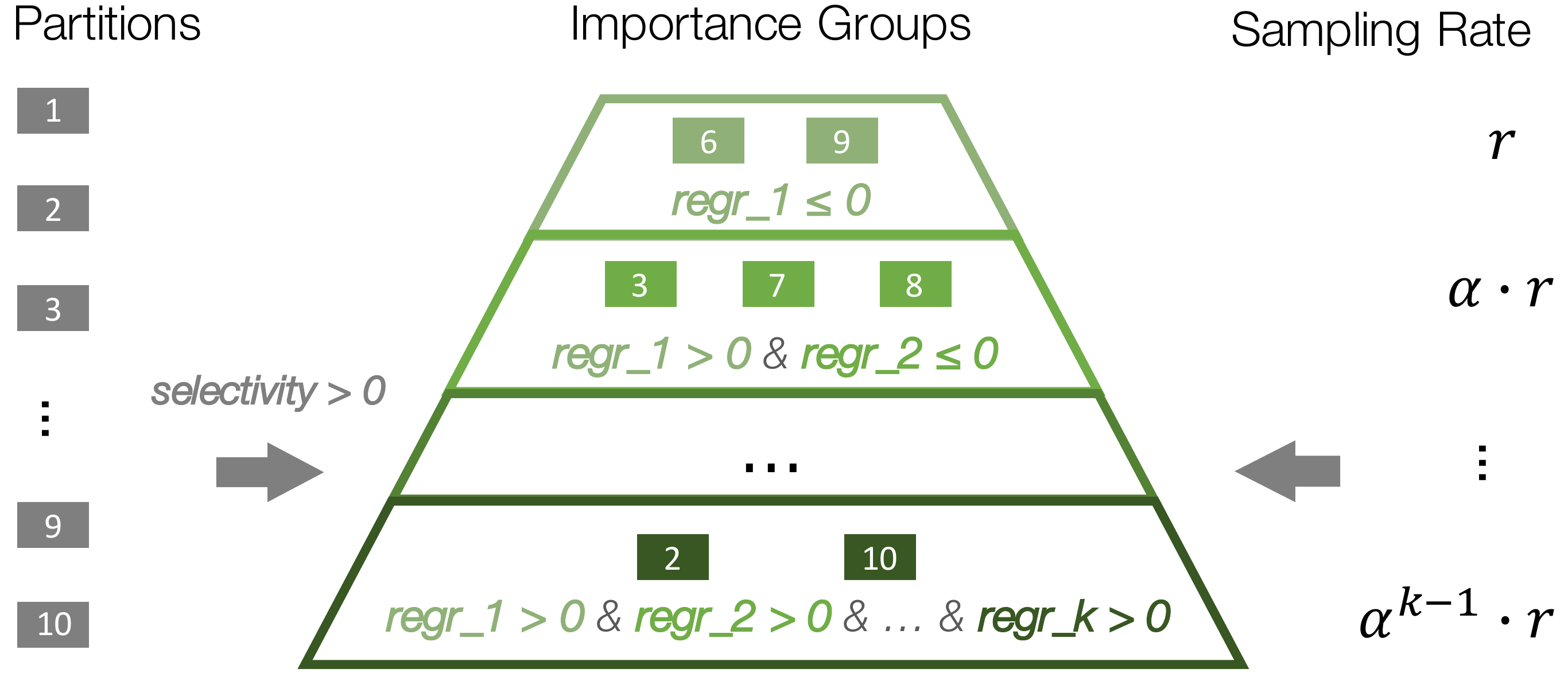}
\caption{The trained regressors are used to classify input partitions into groups of different importance. The sampling rate decreases by a factor of $\alpha > 1$ from the $i^{th}$ important to the $(i+1)^{th}$ important group.}
\label{fig:tier}
\vspace{-1em}
\end{figure}

The feature vectors can help assess partition contribution. Consider the query:
\texttt{SELECT SUM(X), Y FROM table WHERE\\ Z > 1 GROUP BY Y}.
The subset of partitions that answer this query well should contain large values of \texttt{X}, many rows that satisfy the predicate and many distinct values of \texttt{Y}. Feature vectors are correlated with these desired properties: measure statistics (e.g. max, std) can help reveal large values of \texttt{X}, selectivity measures the fraction of the partition that is relevant to the query, and heavy hitter and distinct value statistics summarize the distribution of groups. However, it is challenging to manually quantify how much each feature matters for each query. In our example, it is unclear whether a partition with a high variance of \texttt{X} but few rows that match the predicate should be prioritized over a partition with low variance and many rows that match the predicate.

While it is not obvious how to manually design rules that relate feature vectors to partition contribution, a model may \emph{learn} to do so from examples. An intuitive design is to use partition features as inputs and predict partition weights as outputs, which turns out to be a non-traditional regression problem. The goal of the regressor is to assign a weight vector to $N$ partitions such that the weighted partition choice produces a small approximation error. Given a sampling budget of $n$ partitions, there are exponentially many choices of subsets of partitions of size $n$ and the optimal choice is discontinuous on $n$\footnote{A small change in n can completely change the set of partitions to pick and the weights to assign to them.}. In addition, the decision depends jointly on the {\em set} of partitions chosen; the weight assigned to one partition, for example, may depend on how many other partitions with nearly identical content are picked in the sample. Therefore, a simple, per partition regressor is unable to capture the combinatorial nature of the decision space. Existing solutions~\cite{comb2, comb1} would require significantly more resources and we pursue a lightweight alternative instead.

Given the challenges to directly use learned models to predict sampling probabilities, we propose a design that utilizes the models indirectly for sample size allocation; similar observations were made for using learned models to improve row-level sampling designs for count queries~\cite{lss}. We consider classifying partitions based on their \emph{relative importance} to the query answer into a few importance groups, and apply multiplicatively increasing sampling probability to the more important groups. We detail each of these steps next.

\minihead{Partition Contribution} We consider the ``contribution" of a partition to the answer of a query as its largest relative contribution to any group and any aggregate in the answer. Recall that we denote the aggregates for group $g \in G$ as ${\bf A_{g}} \in \mathbbm{R}^d$, where $d$ is the number of the aggregate functions, and the aggregates for group $g$ on partition $i$ are denoted as ${\bf A_{g, i}} \in \mathbbm{R}^d$. Partition $i$'s contribution is defined as: $\max_{g \in G}\max_{j=1}^d(\frac{{\bf A_{g,i}}[j]}{{\bf A_g[j]}})$. There are several alternative definitions of contribution, such as using the average instead of the max of the ratios, or using absolute values instead of the relatives. Among all variants, the max of the relatives is perhaps the most generous: it recognizes a partition's importance if it helps with \emph{any} aggregates in \emph{any} groups, and is not biased towards large groups or aggregates with large absolute values. We find that our simple definition above already leads to good empirically results.

\minihead{Training} Given the partition contributions for all queries in the training data, we train a set of $k$ models to distinguish the relative importance of partitions. When $k$ is large, training the set of models is equivalent to solving the regression problem in which we are directly predicting partition contribution from the feature vector; when $k$ is small, the training reduces to a simpler multiclass classification problem. The $k$ models discretize partition contribution into $k+1$ bins, and we choose exponentially spaced bin boundaries: the number of partitions that satisfy the $i^{th}$model increase exponentially from the number of partitions that satisfy the $(i+1)^{th}$ model. In particular, the first model identifies all partitions that have non zero contribution to the query and the last model identifies partitions whose contribution is ranked in the top 1\% of all partitions\footnote{The small number of positive examples make it challenging to train an accurate model beyond 1\%.}. We use the \texttt{XGBoost} regressor as our base model, and provide additional details of the training in Appendix B.2.

\minihead{Testing} During test time, we run partitions through a funnel that utilizes the set of trained models as filters and sort partitions into different importance groups (\autoref{fig:tier}). The advantage of building a funnel is that it requires partitions to pass more filters as they advance to the more important groups, which help limit the impact of inaccurate models. We list the procedure in Algorithm~\ref{alg:cascade}. We start from all partitions with non zero \texttt{selectivity\_upper} feature; as discussed in \S~\ref{sec:stats}, this filter has perfect recall but varying precision depending on the complexity of the predicates. We run the partitions through the first trained model, and move the ones that pass the model to the next stage in the funnel. We repeat this process, each time taking the partitions at the end of the funnel, running them through a more restrictive filter (model) and advance ones that pass the filter into the next stage until we run out of filters. 

\begin{algorithm}[t]
\caption{Group partitions by importance.}
\label{alg:cascade}
\begin{algorithmic}[1]
\Require{partition features  $F$}
\Function{ImportanceGroup}{$F$, {\sf parts, regressors}} 
\State {\sf groups}.add(\textproc{FilterByPredicate}($F$, {\sf parts})) 
\For {{\sf regr} $\in$ {\sf regressors}}
\State {\sf to\_examine} $\gets$ {\sf groups[-1]}
\State {\sf to\_pick} $\gets$  {\sf p} $\in$ {\sf to\_examine} s.t. {\sf regr(}$F[p]$) $> 0$
\State {\sf groups[-1]} $\gets $  {\sf to\_examine}.difference({\sf to\_pick})
\State {\sf groups}.add({\sf to\_pick})
\EndFor
\State \Return {\sf groups}
\EndFunction
\end{algorithmic}
\end{algorithm}

We then split the sampling budget such that more important groups get a greater proportion of the budget. We implement a sampling rate that decays by a factor of $\alpha > 1$ from the $i^{th}$ important to the $(i+1)^{th}$ important group. We investigate the impact of the decay rate $\alpha$ in the sensitivity analysis (Appendix C.2). In general, increasing $\alpha$ improves the overall performance especially when the trained models are accurate, but the marginal benefit decreases as $\alpha$ becomes larger. If the trained models are completely random however, a larger $\alpha$ would increase the variance of the estimate. We have found that a decay rate of $\alpha=2$ with $k=4$ models works well across a range of datasets and layouts empirically. However, it is possible to fine-tune $\alpha$ for each dataset to further improve the performance and we leave the fine-tuning to future work.

\subsection{Outliers}
\label{sec:outlier}
Finally, we observe that datasets often exhibit significant skew in practice (example in \S~\ref{sec:intro}). Prior work in AQP has shown that augmenting random samples with a small number of samples with outlying values or from rare groups helps reduce error caused by the skewness~\cite{smallgroup,outlier}. We recognize the importance of handling outliers and allocate a small portion of the sampling budget for outlying partitions. 

We are especially interested in partitions that contain a rare distribution of groups for \texttt{GROUP BY} queries. These partitions are not representative of other partitions and should be excluded from clustering. To identify such partitions, we take advantage of the occurrence bitmap feature that tracks which heavy hitters are present in a partition. We put partitions with identical bitmap features for columns in the \texttt{GROUP BY} clause in the same group and consider a bitmap feature group outlying if its size is small both in absolute ($<10$ partitions) and relative terms ($<10$\% the size of the largest group). For example, if there are 100 such bitmap feature groups and 10 partitions per group, we do not consider any group as outlying although the absolute size of each group is small. We allocate up to 10\% of the sampling budget to evaluate outliers. We have empirically found that increasing the outlier budget further does not significantly improve the performance using only the outliers we consider. Exploring alternative ways to identify outliers could be an interesting area for improvement for future works. 

\section{Evaluation}
In this section, we evaluate the empirical performance of \sysname. Experiments show that: 
\begin{enumerate}[topsep=1pt,itemsep=0.5pt,parsep=0.5pt]
\item \sysname consistently outperforms alternatives on a variety of real-world datasets, delivering $2.7-70\times$ reduction of data read to achieve the same average relative error compared to uniform partition sampling, with storage overhead ranging from 12KB to 103KB per partition.
\item Every component of \sysname and every type of features contribute meaningfully to the final performance.
\item \sysname works across datasets, partitioning schemes, partition counts and generalizes to unseen queries. 
\end{enumerate}

\subsection{Methodology}
In this subsection, we describe the experimental methodology, which includes the datasets, query generation, methods of comparison and error metrics. 
\newpage
\subsubsection{Datasets}
\label{sec:ds}
We evaluate on four real-world datasets that are summarized below. We include a specification of the table schema in Appendix A. 

\minihead{TPC-H*} \textcolor{black}{Data is generated from a Zipfian distribution with skewness of 1 and a scale factor of 1000~\cite{datagen}. We denormalize all tables against the $lineitem$ table. The resulting table has 6B rows, with 14 numeric columns and 31 categorical columns.} Data is sorted by column \texttt{L\_SHIPDATE}.

\minihead{TPC-DS*}  $catalog\_sales$ table with a scale factor of 1 from TPC-DS, joined with dimensions tables $item$, $date\_dim$, $promotion$ and $customer\_demographics$, with 4.3M rows, 21 numeric columns and 20 categorical columns. Data is sorted by columns \texttt{year}, \texttt{month} and \texttt{day}.

\minihead{Aria} Production service request log at Microsoft with 10M rows, 7 numeric columns and 4 categorical columns~\cite{diff,storyboard}. Data is sorted by categorical column \texttt{TenantId}. 

\minihead{KDD} KDD Cup'99 dataset on network intrusion detection with 4.8M rows, 27 numeric columns and 14 categorical columns~\cite{kdd}. Data is sorted by numeric column \texttt{count}.\\\\
By default we use a partition count of 1000, the smallest size from which partition elimination becomes interesting. The \texttt{TPC-H*} dataset (sf=1000) has 2844 partitions, with a partition size of about 2.5GB, consistent with the scale of the big-data workloads seen in practice. In the sensitivity analysis, we further investigate the effect of the partition count (\S~\ref{sec:block}), and data layouts on the performance (\S~\ref{sec:dl}).

\subsubsection{Query Set}
\label{sec:qgen}
To train \sysname, we construct a training set of 400 queries for each dataset by sampling at random the following aspects: 
\begin{itemize} [topsep=0.5pt,itemsep=0.5pt,parsep=0.5pt]
\item between 0 and 8 columns as the group-by columns 
\item between 0 and 5 predicate clauses; each of which picks a column, an operator and a constant at random
\item between 1 and 3 aggregates over one or more columns  
\end{itemize}

We generate a held-out set of 100 test queries in a similar way. For \texttt{TPC-H*}, we include an additional test set of 10 TPC-H queries (\S~\ref{sec:gen}). We ensure that there are no identical queries between the test and training sets and that there is substantial entropy in our choice of predicates, aggregates and grouping columns.

\begin{figure*}[ht]
\centering
\includegraphics[width=\textwidth]{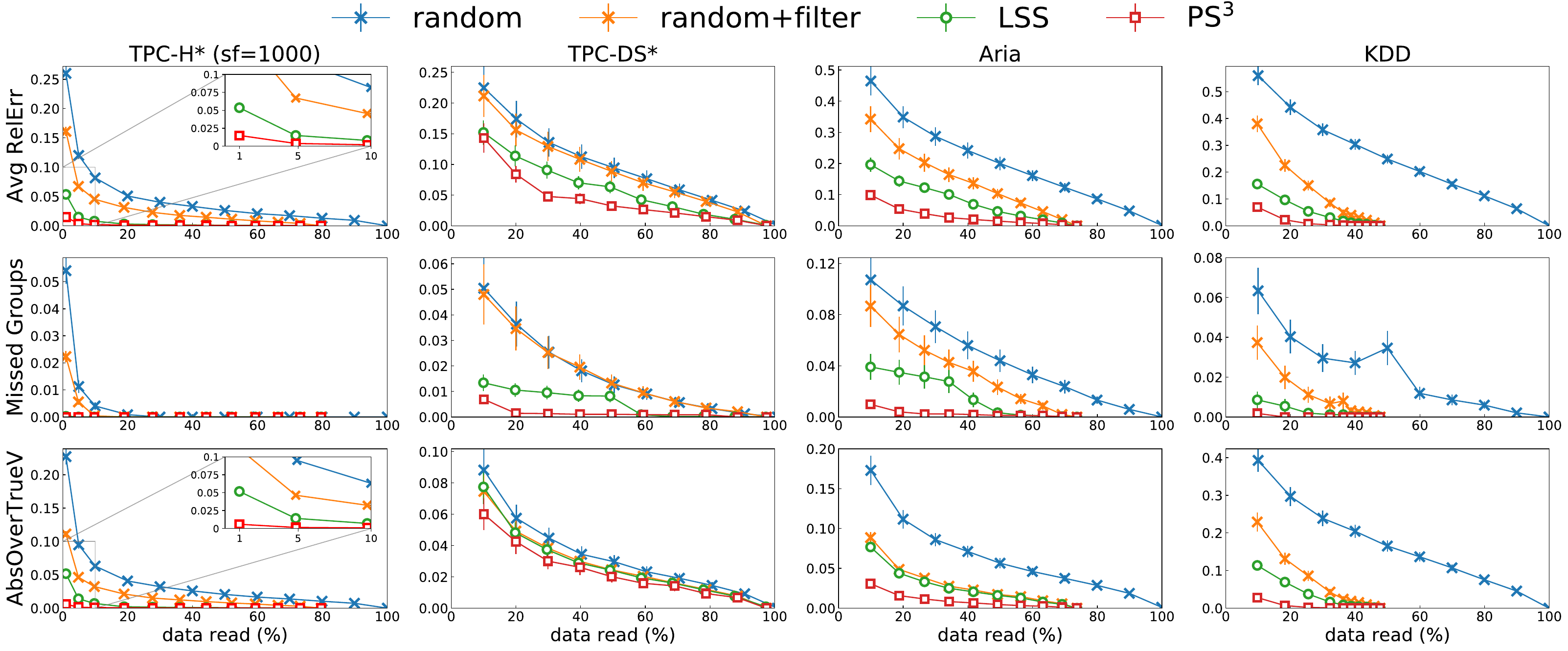}
\caption{\textcolor{black}{Comparison of error under varying sampling budget on four datasets, lower is better. \sysname (red) consistently outperforms others across datasets and different error metrics.} }
\label{fig:main}
\vspace{-0.5em}
\end{figure*}

\subsubsection{Methods of Comparison}
All methods except for simple random sampling have access to feature vectors, and use the \texttt{selectivity\_upper} feature to filter out partitions that do not satisfy the predicate before sampling. Recall that this filter has false positives but no false negatives. All methods have access to the same set of features. We report the average of 10 runs for methods that use random sampling. 

\minihead{Random Sampling} Partitions are sampled uniformly at random. Aggregates in the answer are scaled up by the sampling rate.

\minihead{Random+Filter} Same as random sampling except that only partitions that pass the selectivity filter are sampled. This is only achievable with the use of summary statistics. 

\minihead{Learned Stratified Sampling (LSS)} A baseline inspired by prior work on learned row-level stratified sampling~\cite{lss}. We rank partitions by the model's prediction and perform stratification such that each strata covers partitions whose predictions fall into a consecutive range. We made three modifications to LSS to enable partition-level sampling: moving training from online to offline for I/O savings, changing inputs and outputs to operate on partitions instead of rows, and adopting a different stratification strategy. We include a detailed description of the modifications in Appendix C.1.

\minihead{\sysname} A prototype that matches the description given so far. Unless otherwise specified, default parameter values for \sysname in all experiments are $k=4, \alpha=2$ and up to a 10\% sampling budget dedicated to outliers.

\subsubsection{Error Metric}
Similar to prior work~\cite{congressional, smallgroup, experiences}, we report multiple accuracy metrics. It is possible, for example, for a method to have a small absolute error but miss all small groups and small aggregate values. We therefore consider all three metrics below for a complete picture. 

\minihead{Missed Groups} Percentage of groups in the true answer that are missed by the estimate. 

\minihead{Average Relative Error} The average of the relative error for each aggregate in each group. For missed groups, the relative error is counted as 1. 

\minihead{Absolute Error over True} The average absolute error value of an aggregate across groups divide by the average true value of the aggregate across groups, averaged over multiple aggregates.

\subsection{Macro-benchmarks}
\label{sec:main}
We compare the performance of methods of interest under varying sampling budgets on four datasets (\autoref{fig:main}). The closer the curve is to the bottom left, the better the results.

While the scale of the three error metrics is different, the ordering of the methods is relatively stable. Using the selectivity feature to filter out partitions that do not satisfy the predicate strictly improves the performance for all methods, except on datasets like \texttt{TPC-DS*} where most partitions pass the predicate. The modified LSS (green) clearly improves upon random sampling by leveraging the correlation between feature vectors and partition contribution, consistent with findings of prior work.

Overall, \sysname consistently outperforms alternatives across datasets and error metrics. On our large scale experiment with the \texttt{TPC-H*} data, \sysname achieves an average relative error of 1.5\% with a 1\% sampling rate. With a 1\% same sampling rate, \sysname improves the error achieved by $17.5\times$ compared to random sampling, $10.8\times$ compared to random sampling with filter and $3.6\times$ compared to LSS (read from intersections between the baseline curves and a vertical line at 1\% sampling rate). To achieve an average relative error of 1.5\%, \sysname reduces the fraction of data read by over $70\times$ compared to random sampling, over $40\times$ compared to random sampling with filter and $5\times$ compared to LSS (read from the intersections between baseline curves and a horizontal line at 1.5\% error rate). We observe similar trends on the three smaller datasets but the performance gap is smaller: \sysname reduces the data read by $2.7\times$ to $8.5\times$ compared to simple random sampling to achieve $\leq 10\%$ average relative error.

\begin{table}
\centering
\small
\caption{\textcolor{black}{Average speedups for query latency and total compute time under difference sampling rates on the \texttt{TPC-H*} dataset. } }
\begin{tabular}{lcccc}
\toprule
 & 1\% & 5\% & 10\% & 100\%\\
 \midrule
Query Latency & 4.7$\times$ & 1.6$\times$ & 1.5$\times$ & - \\
Total Compute Time & 105.3$\times$& 19.6$\times$ & 11.4$\times$ & -\\
 \bottomrule
\end{tabular}
\label{tab:time}
\vspace{-1em}
\end{table}

We additionally show that the fraction of data read is a reliable proxy for reductions in resources used, measured by total compute time. We evaluate example queries on the \texttt{TPC-H*} dataset using SCOPE clusters~\cite{scope, scope2}, Microsoft's main batch analytics platform, which consist of tens of thousands of nodes. Table~\ref{tab:time} shows that reading 1\%, 5\% and 10\% of the partitions results in a near linear speedup of 105.3$\times$, 19.6$\times$, 11.4$\times$ in the total compute time. Improvement of query latency however, is less than linear and depends on stragglers and other concurrent jobs on the cluster.

\subsection{Overheads}
\begin{table}
\centering
\small
\caption{Per partition storage overhead of the summary statistics (in KB) for each dataset.}
\begin{tabular}{lrrrrr}
\toprule
Dataset & Total  & Histogram & HH & AKMV & Measure \\
 \midrule
 \texttt{TPC-H*} & 84.25& 9.52 & 13.26 & 55.31 & 6.16 \\
 \texttt{TPC-DS*}  & 103.49 & 10.51 & 4.67 & 81.45 & 6.86 \\
\texttt{Aria}  & 18.38 & 1.42 & 0.81 & 15.19 & 0.97 \\
\texttt{KDD}   &  12.00 & 2.19& 0.82& 5.29 & 3.70 \\
\bottomrule
\end{tabular}
\label{tab:sapce}
\vspace{-1em}
\end{table}

We report the space overhead of storing summary statistics in ~\autoref{tab:sapce}. The statistics are computed for each column and therefore require a constant storage overhead per partition. The overheads range from 12KB to 103KB across the four datasets. The larger the partition size, the lower the relative storage overhead of the statistics. \textcolor{black}{For example, with a partition size of 2.5GB, the storage overhead is below 0.003\% for the \texttt{TPC-H*} dataset.} 

The AKMV sketch for estimating distinct values takes the most space compared to other sketches. If the number of distinct values in a column is larger than $k$ (we use $k=128$), the sketch has a fixed size; otherwise the sketch size is proportional to the number of distinct values. The \texttt{KDD} dataset, for example, has more columns but a smaller AKMV sketch size compared to the \texttt{Aria} dataset since a number of its columns are binary. 

We also report the single-thread latency of the partition picker (Algorithm~\ref{alg:test}) in Table~\ref{tab:latency}, measured on an Intel Xeon E5-2690 v4 CPU. Our prototype picker is implemented in Python using the \texttt{XGBoost} and \texttt{Sklearn} libraries. Overall, the overhead is a small fraction of the query time, ranging from 86.5ms to around 1s across datasets. In comparison, the average query takes tens of total computation hours on the \texttt{TPC-H*} dataset. As the number of partitions and the dimension of the feature vectors increase, the total overhead increases and the clustering component takes up an increasing proportion of the overhead. The overhead can be further reduced via optimization such as performing clustering in parallel across different importance groups.  

\begin{table}
\centering
\small
\caption{\textcolor{black}{Range of the average picker overhead across sampling budgets for each dataset (in milliseconds).} }
\begin{tabular}{l@{\hskip7pt}r@{\hskip7pt}r@{\hskip7pt}r@{\hskip7pt}r}
\toprule
 & \texttt{Aria} & \texttt{KDD} & \texttt{TPC-DS*} & \texttt{TPC-H*}\\
 \midrule
 Total & 89.9$\pm$4.7  & 106.4$\pm$4.9 &  219.6$\pm$4.7  & 1002.1$\pm$13.3 \\ 
 Clustering & 24.1$\pm$5.0   & 58.0$\pm$2.2 &148.0$\pm$5.4  & 802.4$\pm$12.8 \\
 \bottomrule
\end{tabular}
\label{tab:latency}
\vspace{-0.5em}
\end{table}

\subsection{Lesion Study}
In this section, we take a closer look at individual components of the picker and their impact on the final performance, as well as the importance of partition features. 

\subsubsection{Picker Lesion Study}
\begin{figure}
\centering
\includegraphics[width=0.9\columnwidth]{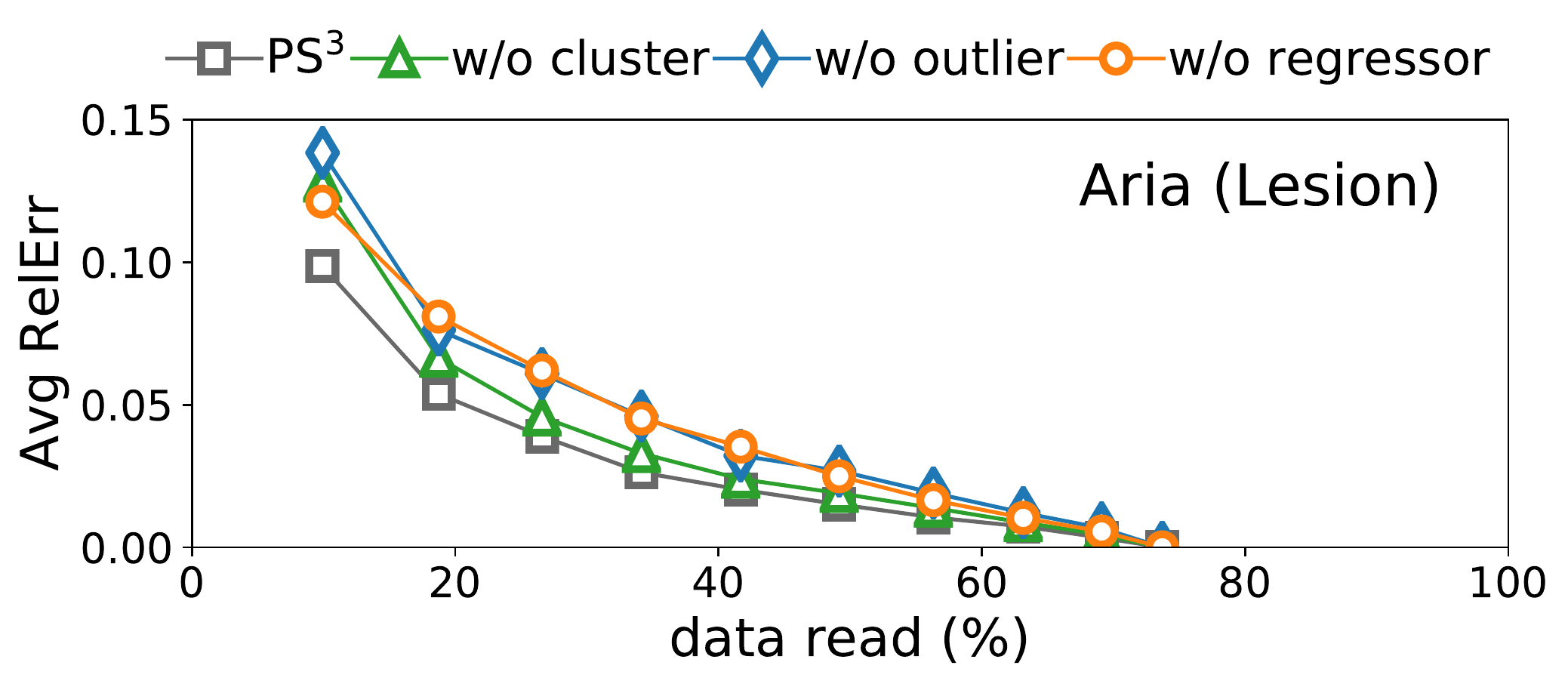}
\includegraphics[width=0.9\columnwidth]{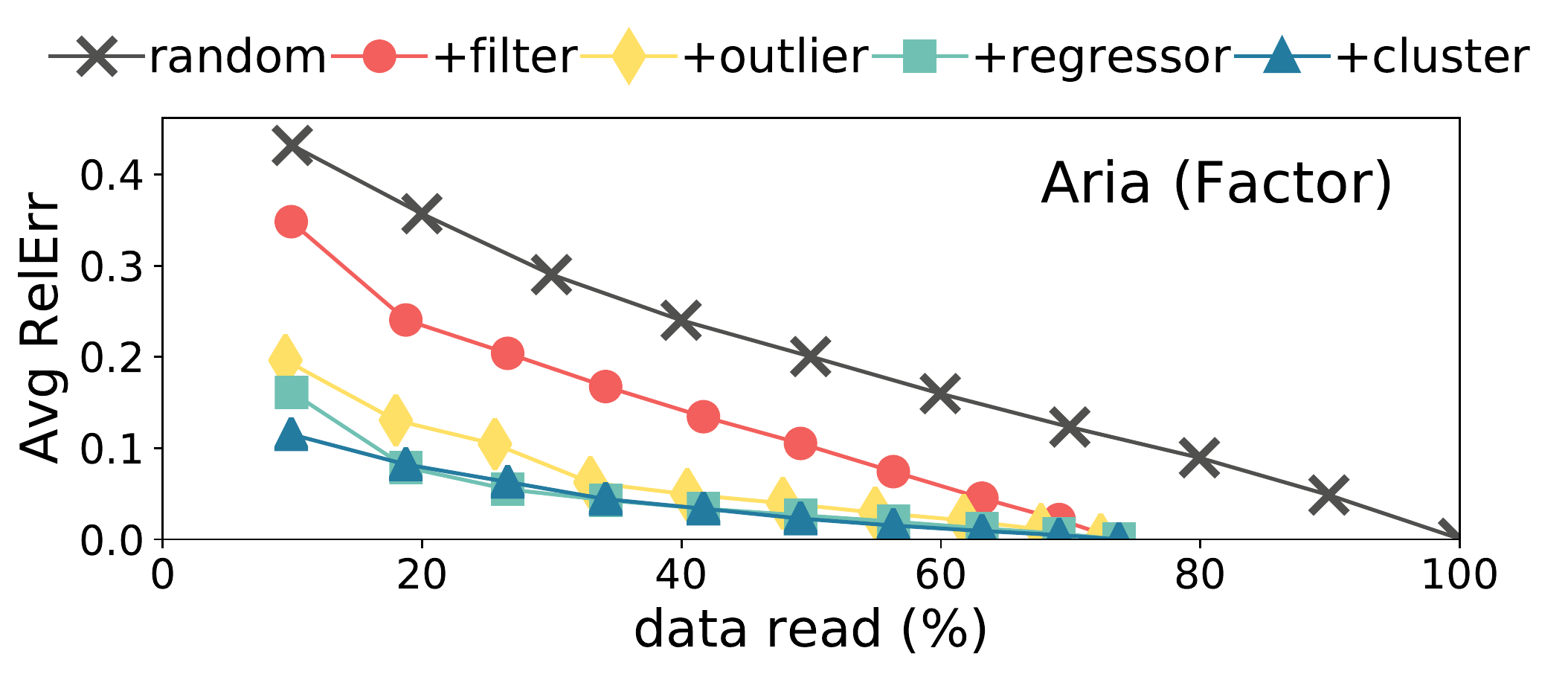}
\caption{Lesion study and factor analysis on the Aria dataset. Each component of our system contributes meaningfully to the final accuracy. Results are similar on other datasets. }
\label{fig:lesion}
\vspace{-0.5em}
\end{figure}

We inspect how the three components of the partition picker introduced in \S~\ref{sec:algo} impact the final accuracy. To examine the degree to which a single component impacts the performance, we perform a lesion study where we remove each component from the picker while keeping the others enabled (Figure~\ref{fig:lesion}, top). To disable clustering (\S~\ref{sec:cluster}), we use random sampling to select samples. To disable identification of outlier partitions (\S~\ref{sec:outlier}), we take away the sampling budget dedicated to outliers. To disable the regressor (\S~\ref{sec:clf}), we apply the same sampling rate to all partitions. The result shows that the final error increases when each component is disabled, illustrating that each component is necessary to achieve the best performance. 

We additionally measure how the three components contribute to overall performance. Figure~\ref{fig:lesion} (bottom) reports a factor analysis. We start from the simple random sampling baseline (random). Using \texttt{selectivity\_upper} $\geq 0$ as a filter (+filter) strictly improves the performance. \textcolor{black}{Similar to the lesion study, we enable each component on top of the filter (not cumulative) while keeping others disabled. The results show that the identification of outlier partitions (+outlier) contributes the least value individually and the use of clustering (+cluster) contributes the most.}
\begin{figure}
\includegraphics[width=\columnwidth]{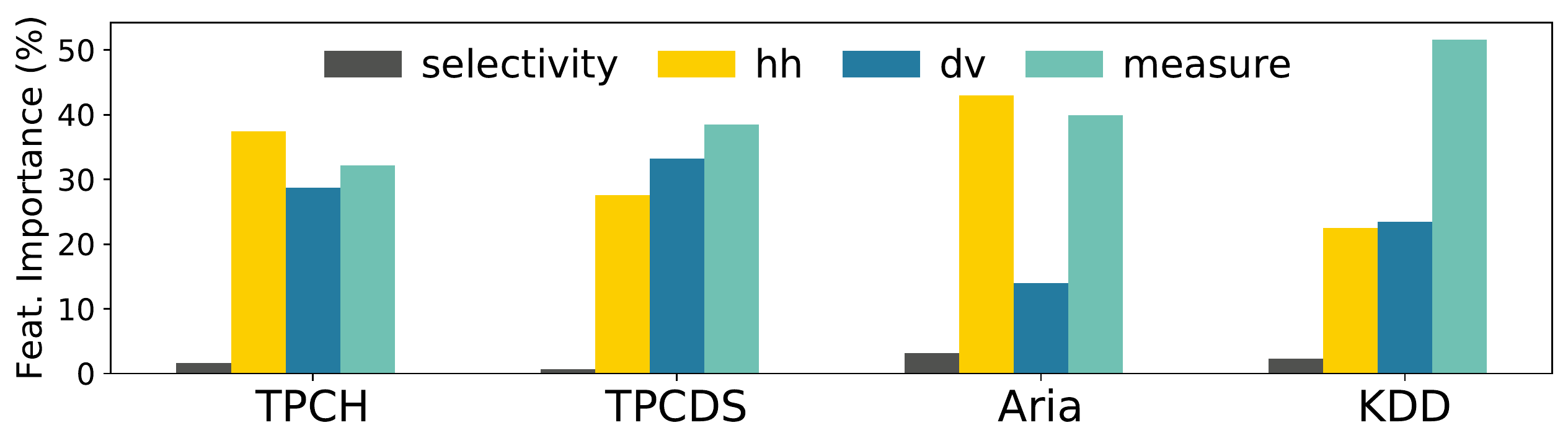}
\caption{Feature importance for the regressors. The higher the percentage, the more important the statistics are to the regressor's accuracy.}
\label{fig:imp}
\vspace{-1em}
\end{figure}

\subsubsection{Feature Importance}
\label{sec:featimp}

We divide partition features into four categories based on the sketches used to generate them: selectivity, heavy hitter, distinct value and measures. We investigate the contribution of features in each component of \sysname. The outlier component depends exclusively on the heavy hitter features. The clustering component uses all four feature types and we report the list of features selected for each dataset in Appendix B.1. For the learned component, we measure the regressors' feature importance via the ``gain" metric, which reports the improvement in accuracy brought by a feature to the branches it is on~\cite{imp}. For each dataset, we report the gain for features in each category as a percentage of the total gain aggregated over all learned models. The larger the percentage, the more important the feature is to the final accuracy. We report the result in \autoref{fig:imp}.

Overall, all four types of features contribute to the regressor accuracy, but the relative importance varies across the datasets. Selectivity estimates, despite being less useful for regressors, are useful to filter out partitions that do not contain any rows satisfying the predicate.

\subsection{Sensitivity Analysis}
\label{sec:sensitivity}
In this section, we evaluate the sensitivity of the system's performance to changes in setups and parameters.
\subsubsection{Effect of Data Layouts}
\label{sec:dl}
\begin{figure}
\centering
\includegraphics[width=0.95\columnwidth]{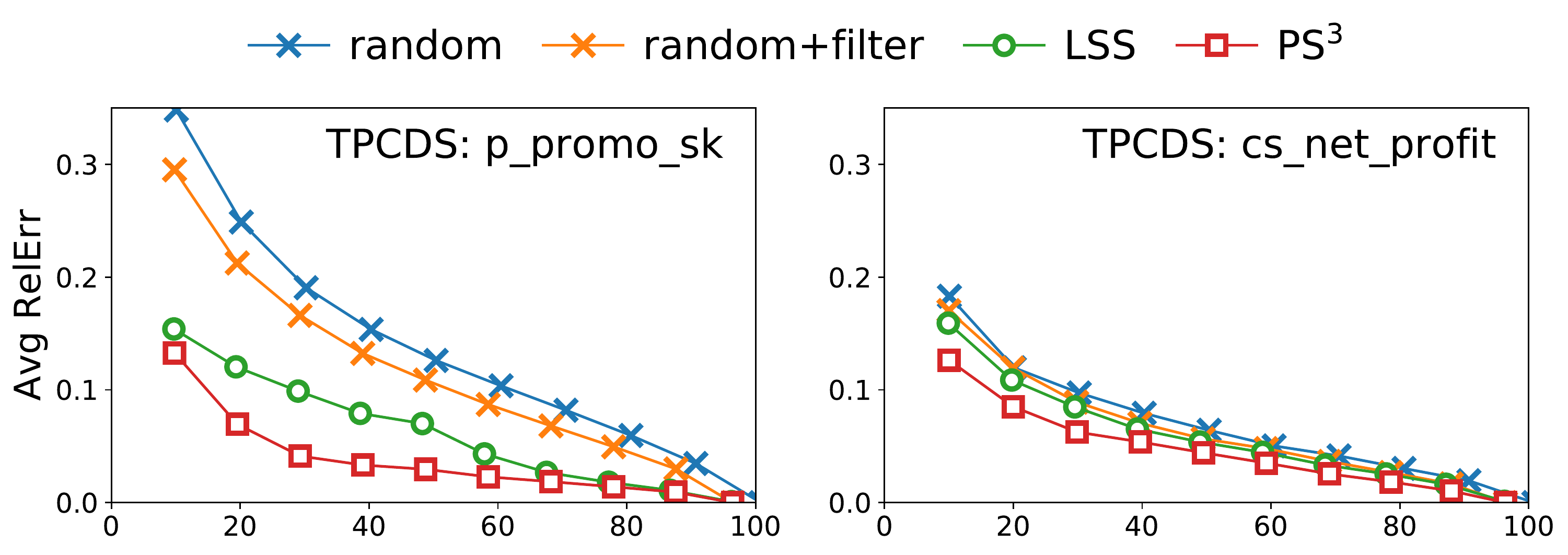}
\includegraphics[width=0.95\columnwidth]{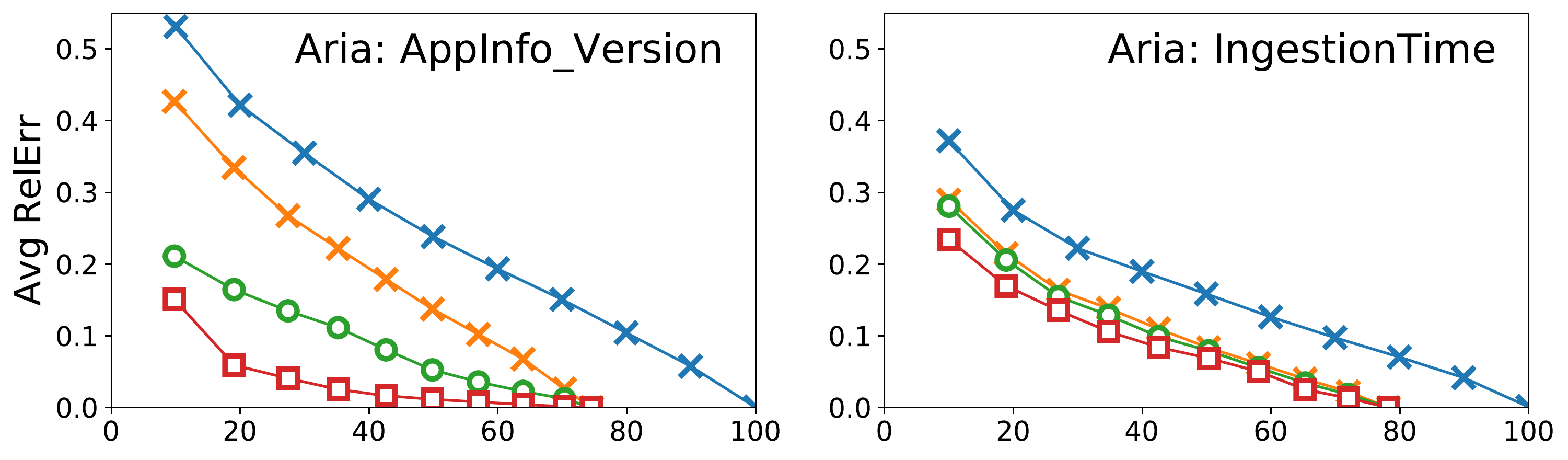}
\includegraphics[width=0.95\columnwidth]{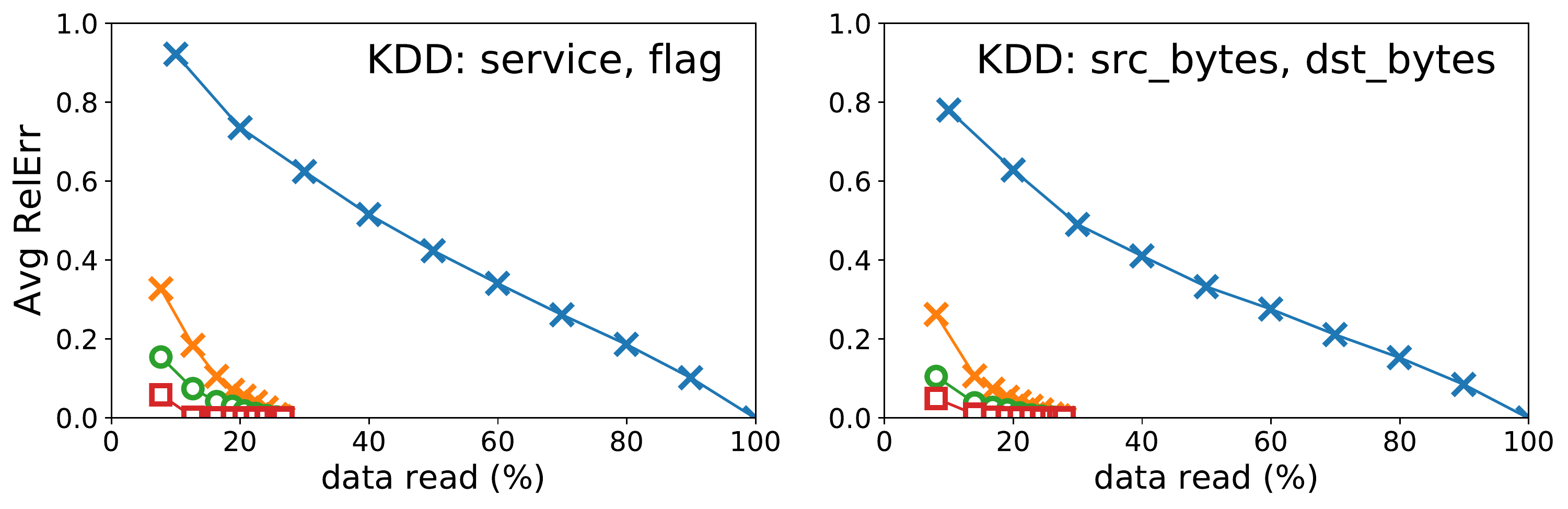}
\caption{Our method consistently outperforms alternatives across datasets and data layouts.}
\label{fig:dl}
\vspace{-0.5em}
\end{figure}

One of our design constraints is to be able to work with data in situ. To assess how \sysname performs on different data layouts, we evaluate on two additional layouts for each dataset using the same training and testing query sets from experiments in \S~\ref{sec:main}. Figure~\ref{fig:dl} summarizes the average relative error achieved under varying sampling budgets for the six combinations of datasets and data layouts.

\sysname consistently outperforms alternatives across the board, but the sizes of the improvements vary across datasets and layouts. Overall, the more random/uniform the data layout is, the less room for improvement for importance-style sampling. For example in the \texttt{TPC-DS*} dataset, the layout sorted by column \texttt{cs\_net\_profit} is more uniform than the layout sorted by column \texttt{p\_promo\_sk}, since random sampling achieves a much smaller error under the same sampling budget in the former layout. LSS is only marginally better than random in the former layout, indicating a weak correlation between features and partition importance. 

As a special case, we explicitly evaluate \sysname on a random layout for the \texttt{TPC-H*} dataset with a scale factor of 1 (Figure~\ref{fig:tpch}, left). As expected, sampling partitions uniformly at random performs well on the random layout. \sysname underperforms random sampling in this setting, but the performance difference is small. Realistically, we do not expect \sysname to be used for random data layouts; users would have chosen random sampling were they paying the cost to maintain a random data layout~\cite{warehouse}.

\subsubsection{Effect of Query Selectivity}
\label{sec:selectivity}
\begin{figure}
\centering
\includegraphics[width=\columnwidth]{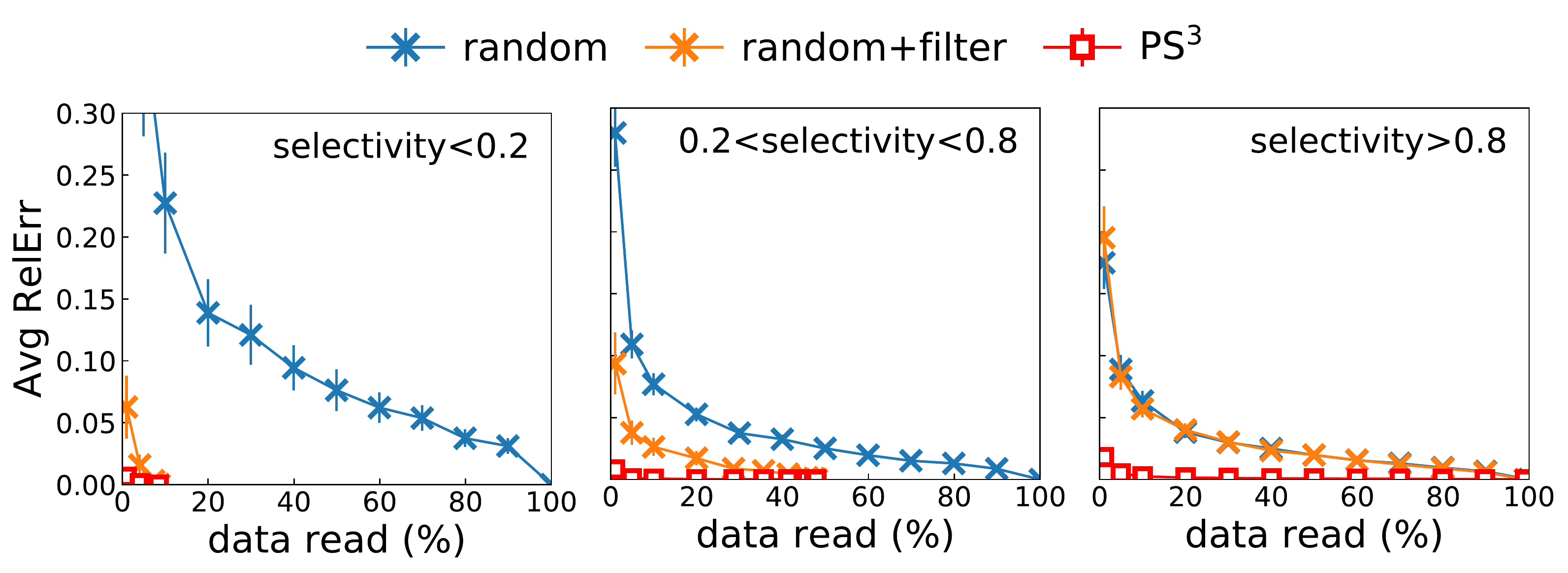}
\caption{\textcolor{black}{Performance breakdown by query selectivity on the \texttt{TPC-H*} dataset (sf=1000).} }
\label{fig:selectivity}
\vspace{-1em}
\end{figure}

We investigate how queries with different sensitivities benefit from \sysname. \autoref{fig:selectivity} reports the error breakdown by query selectivity for random partition-level sampling and \sysname on the \texttt{TPC-H*} dataset; other datasets show similar trends. Compared to naive random partition level sampling (blue), \sysname offers more improvements for more selective queries (selectivity $< 0.2$), since the selectivity feature effectively filters out a large fraction of partitions that are irrelevant to the query. Compared to random partition level sampling with the selectivity filter (orange), \sysname offers more improvements for non-selective queries (selectivity $>0.8$), since they have larger errors at small sampling rates.

\subsubsection{Effect of Partition Count}
\label{sec:block}
\begin{figure}
\centering
\includegraphics[width=\columnwidth]{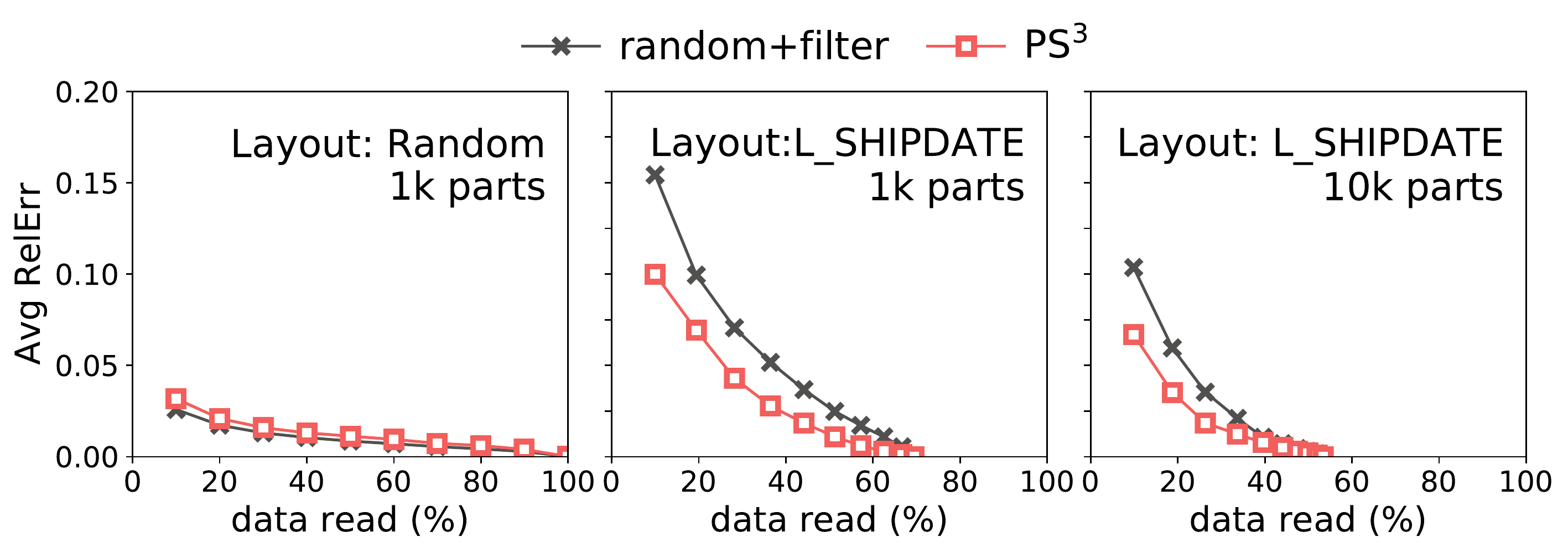}
\caption{Comparison of \texttt{TPC-H*} (sf=1) results on different data layouts and total number of partitions.}
\label{fig:tpch}
\vspace{-0.5em}
\end{figure}

In this subsection, we investigate the impact of partition count on the final performance. We report results on the \texttt{TPC-H*} dataset (sf=1) with 1000 and 10,000 partitions in the middle and right plot of ~\autoref{fig:tpch}. Compared to results on the same dataset with fewer partitions, the percentage of partitions that can be skipped increases with the increase of the number of partitions. In addition, as the partition count increases, the error achieved under the same sampling fraction becomes smaller. However, the overheads of \sysname also increase with the number of partitions. Specifically, the storage overhead for per-partition statistics increases linearly with the number of partitions. The latency of the partition picker also increases with the partition count. Perhaps more concerning is the increase in I/O costs. The larger the partition count, the smaller the size of each partition. In the limit when each partition only contains one row, partition-level sampling is equivalent to row-level sampling, which is expensive to construct as discussed earlier (\S~\ref{sec:intro}).  

\subsubsection{Generalization Test on TPC-H Queries}
\label{sec:gen}
\begin{figure}
\centering
\includegraphics[width=\columnwidth]{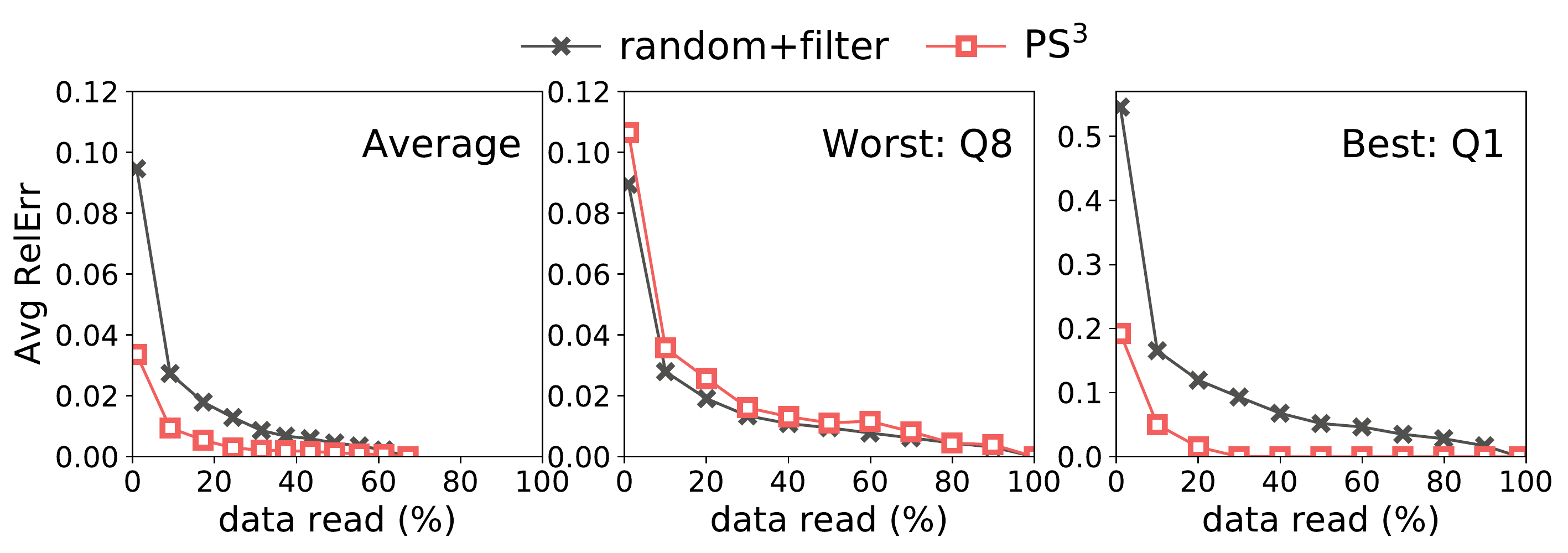}
\caption{\textcolor{black}{The average, worst and best results from a generalization test on unseen TPC-H queries (sf=1).} }
\label{fig:gen}
\vspace{-0.5em}
\end{figure}
To further assess the ability of the trained models to generalize to unseen queries, we test \sysname trained on the randomly generated training queries with TPC-H schema (described in ~\S~\ref{sec:qgen}) on 10 unseen TPC-H queries supported by our query scope\footnote{Q4 is excluded since it operates on the \emph{orders} table.}; the set of aggregate functions and group by columnsets are shared between the train and test set. We generate 20 random test queries for each TPC-H query template. We report the average, worst and best performances across the test queries on the \texttt{TPC-H*} dataset (sf=1000) in ~\autoref{fig:gen}. \textcolor{black}{On average, \sysname is still able to outperform uniform partition sampling, despite the larger domain gap between training and test set compared to experiments conducted in \S~\ref{sec:main}. We report a detailed performance breakdown in Appendix C. Overall, we enable larger improvements on queries such as Q1, where a small number of partitions contain rare groups or outlying aggregate values. Our improvements are limited on Q8 which has a more complex aggregate and a nested query.} 

\begin{table}[t]
\centering
\small
\caption{AUC for different clustering algorithms; smaller is better.}
\begin{tabular}{l|r|r|r}
\toprule
 & HAC(single) & HAC(ward) & KMeans \\
 \midrule
 \texttt{TPCDS}  & 12.1 & 4.2  &4.2 \\
\texttt{Aria}  & 3.2 & 2.6 & 2.7 \\
\texttt{KDD}   & .71 & .58  & .55 \\
\bottomrule
\end{tabular}
\label{tab:cluster}
\vspace{-0.5em}
\end{table}

\subsubsection{Effect of Clustering Algorithm}
\label{sec:evalcluster}
We evaluate the effect of clustering algorithm choice on the clustering performance. 

We compare a bottom-up clustering algorithm (Hierarchical Agglomerative Clustering, or HAC) to a top-down algorithm (KMeans). For HAC, we evaluate two linkage metrics: the ``single" linkage minimizes the minimum distances between all points of the two merged clusters, while the ``ward" linkage minimizes the variances of two merged clusters. For each dataset, we evaluate the \emph{average relative error} for estimating the query answer, and report the area under the error curve under different sampling budgets (\autoref{tab:cluster}). The smaller the area, the better the clustering performance. We also include results from a similar evaluation on the impact of the feature selection procedure in Appendix B.1. 

HAC using the ``ward" linkage metric and K-Means consistently produce similar results, suggesting that the clustering performance is not dependent on the choice of the clustering algorithm. The single linkage metric, however, produces worse results especially on the \texttt{TPCDS} dataset.

\section{Related Work}
\begin{sloppypar}
In this section, we discuss related work in sampling-based AQP, data skipping, and partition-level sampling. 

\minihead{Sampling-based AQP} Sampling-based approximate query processing has been extensively studied, where query results are estimated from a carefully chosen subset of the data~\cite{bullet}. A number of techniques have been proposed to improve upon simple row-level random sampling, for example, by using auxiliary data structures such as an outlier index~\cite{outlier} or by biasing the samples towards small groups~\cite{aqua,congressional}. Prior work has shown that, despite the improvements in sampling techniques, it is often difficult to construct a sample pool that offers good results for arbitrary queries given feasible storage budgets~\cite{quickr}. Instead of computing and storing samples apriori~\cite{blinkdb, smallgroup, chaudhuri2001robust}, our work makes sampling decisions exclusively during query optimization.

Prior works have used learning to improve the sampling efficiency for AQP. One line of work uses learning to model the dataset and reduce the number of samples required to answer queries~\cite{deshpande2004model}. Similarly, prior work tries to learn the underlying distribution that generates the dataset from queries, and relies on the learned model to reduce sample size over time~\cite{dblearning}. Our work is closer to works that use learned models to improve the design of the sampling scheme. Recent work proposes a learned stratified sampling scheme wherein the model predictions are used as stratification criteria~\cite{lss}. However, the work focuses on row-level samples and on count queries; we support a broader scope of queries with aggregates and group bys and work with partition-level samples. In the evaluation, we compare against a scheme inspired by learned stratified sampling.  

\minihead{Data Skipping} Our work is also closely related to prior works on data skipping which studied the problems of optimizing data layouts~\cite{skipping, skipping2, partitioning, qdtree} and indexing~\cite{cracking, cracking1, cracking2}, improving data skipping given a query workload. Building on the observation that it is often difficult, if not impossible, to find a data layout that offers optimal data skipping for all queries, we instead choose to work with data in situ. Researchers and practitioners have also looked at ways to use statistics and metadata to prune partitions that are irrelevant to the query. The proposed approaches range from using simple statistics such as min and max to check for predicate ranges~\cite{impala, mysql}, to deriving complex pruning rules for queries with joins~\cite{dip}. Our work is directly inspired by this line of work and extends deterministic pruning to probabilistic partition selection. 

\minihead{Partition-level sampling} Researchers have long recognized the I/O benefits of partition-level sampling over row-level sampling~\cite{clustersampling, seshadri1992sampling}. Partition-level samples have been used to build statistics such as histograms and distinct value estimates for query optimizers~\cite{chaudhuri2004effective, chaudhuri1998random}. Prior work has studied combining row-level and partition-level Bernoulli style sampling for \texttt{SUM, COUNT}, and \texttt{AVG} queries, in which one can adjust the overall sampling rate but each sample is treated equally~\cite{bilevel}. Our work more closely resembles importance sampling where we sample more important partitions with higher probability.

Partition level sampling is also studied in the context of online aggregation (OLA) where query estimates can be progressively refined as more data gets processed, and users can stop the processing when the answer reaches target accuracy~\cite{ci2015efficient,olamr, cola}. Classic work in OLA assume that tuples are processed in a random order, which often require random shuffling as an expensive processing step~\cite{warehouse}. Our approach does not require random layout, and in fact, should not be used if the data layout is random. Prior work has also studied OLA over raw data, which requires an expensive tuple extraction step to process raw files~\cite{olabi}. \sysname can work with data stored in any format as long as per-partition statistics are available and focuses on selecting fewer partitions instead of stopping processing early within a partition, since the most expensive operation for our setup is the I/O cost of reading the partition.
\end{sloppypar}

\balance


\section{Discussion and Future Work}
\label{sec:future}
Our work shows promise as a first step towards using learning techniques to improve upon uniform partition-level sampling. We highlight a few important areas for future work below. 

First, our system is designed mainly for read-only and append-only data stores, so the proposed set of sketches should be reconsidered if deletions and edits to data must be supported. Furthermore, the partition picker logic must be retrained when the summary statistics of partitions change in a substantial way.

Second, our work only considers generalization to unseen queries in the same workload on the same dataset and data layout. Although retraining can help generalize to unseen columns in the same dataset and layout, supporting broader forms of generalization such as to different data layouts is non-trivial and requires further attention.

Third, our work demonstrates empirical advantages to uniform partition-level sampling on several real-world datasets but provides no apriori error guarantees. Developing error guarantees and diagnostic procedures for failure cases will be of immediate value to practitioners.

\section{Conclusion}
We introduce \sysname, a system that leverages lightweight summary statistics to perform weighted partition selection in big-data clusters. We propose a set of sketches -- measures, heavy hitters, distinct values, and histograms -- to generate partition-level summary statistics that help assess partition similarity and importance. We show that our prototype \sysname provides sizable speed ups compared to random partition selection with a small storage overhead.


\section{Acknowledgement}
\begin{spacing}{0.95}
We thank Laurel Orr for her contributions to an early version of this project. We thank Surajit Chaudhuri and many members of the Stanford InfoLab for their valuable feedback. Kexin Rong, Peter Bailis and Philip Levis were supported in part by affiliate members and other supporters of the Stanford DAWN project---Ant Financial, Facebook, Google, Infosys, NEC, and VMware---as well as Toyota Research Institute, Northrop Grumman, Amazon Web Services, Cisco, the NSF under CAREER grant CNS-1651570 and the NSF under Grant No. CPS-1931750. Any opinions, findings, and conclusions or recommendations expressed in this material are those of the authors and do not necessarily reflect the views of the NSF. Toyota Research Institute ("TRI") provided funds to assist the authors with their research but this article solely reflects the opinions and conclusions of its authors and not TRI or any other Toyota entity.
\end{spacing}

\newpage
\bibliographystyle{abbrv}
\bibliography{ms}  

\section*{APPENDIX}
\setcounter{section}{0}
 \setcounter{subsection}{0}
 \def\thesection{\Alph{section}}
 
 \section{Data Schema}
 \label{app:schema}
 \subsection{TPC-H*} 
We provide the query used to denormalize the \emph{lineitem} table in the \texttt{TPC-H} dataset below. This denormalized table can support 16 out of 22 queries in the TPC-H benchmark (Q1,3,4,5,6,7,8,9,10,12,14,15,17,18,19,21). 
We additionally include two derived columns \texttt{L\_YEAR} and \texttt{O\_YEAR} in the view in order to support group by clauses on these columns (Q7,8,9). Our generalization test (\S~\ref{sec:gen}) includes the following 10 queries: Q1,5,6,7,8,9,12,14,17,18,19. \\
 
 {\footnotesize\noindent
${\tt\bf CREATE\ \ TABLE}$ denorm ${\tt\bf AS}$ \ \ \ \ \\
${\tt\bf SELECT}$
        lineitem.*, customer.*,
   orders.*, part.*, \\
  \hspace*{1em}    partsupp.*, supplier.*, n1.*, n2.*, r1.*, r2.*, \\
    \hspace*{1em} ${\tt\bf datepart}$(yy, o\_orderdate) ${\tt\bf AS}$ o\_year,\\
    \hspace*{1em} ${\tt \bf datepart}$(yy, l\_shipdate) ${\tt\bf AS}$ l\_year\\
${\tt\bf FROM}$ lineitem ${\tt\bf JOIN}$ partsupp ${\tt\bf ON}$ ps\_partkey = l\_partkey ${\tt\bf AND}$ ps\_suppkey = l\_suppkey \\
  \hspace*{1em}   ${\tt\bf JOIN}$ orders ${\tt\bf ON}$ o\_orderkey = l\_orderkey \\
   \hspace*{1em}  ${\tt\bf JOIN}$ part ${\tt\bf ON}$ p\_partkey = ps\_partkey \\
    \hspace*{1em} ${\tt\bf JOIN}$ supplier ${\tt\bf ON}$ s\_suppkey = ps\_suppkey \\
    \hspace*{1em} ${\tt\bf JOIN}$ customer ${\tt\bf ON}$ c\_custkey = o\_custkey \\
    \hspace*{1em} ${\tt\bf JOIN}$ nation ${\tt\bf AS}$ n1 ${\tt\bf ON}$ n1.n\_nationkey = c\_nationkey \\
    \hspace*{1em} ${\tt\bf JOIN}$ nation ${\tt\bf AS}$ n2 ${\tt\bf ON}$ n2.n\_nationkey = s\_nationkey \\
   \hspace*{1em}  ${\tt\bf JOIN}$ region ${\tt\bf AS}$ r1 ${\tt\bf ON}$ r1.r\_regionkey = n1.n\_regionkey \\
   \hspace*{1em}  ${\tt\bf JOIN}$ region ${\tt\bf AS}$ r2 ${\tt\bf ON}$ r2.r\_regionkey = n2.n\_regionkey
}

\subsection{TPC-DS*} 
We provide the query used to denormalize the $catalog\_sales$ table below. The joined dataset contains 4.3M rows, 21 numeric columns and  20 categorical columns. \\

 {\footnotesize\noindent
${\tt\bf CREATE\ \ TABLE}$ denorm\_cs ${\tt\bf AS}$ \ \ \ \ \\
${\tt\bf SELECT}$ catalog\_sales.*, cd.*, item.*, promo.*, date.* \\
    ${\tt\bf FROM}$ catalog\_sales \\
    \hspace*{1em}  ${\tt\bf JOIN}$ item ${\tt\bf ON}$ cs\_item\_sk = i\_item\_sk \\
	\hspace*{1em}  ${\tt\bf JOIN}$ promo ${\tt\bf ON}$ cs\_promo\_sk = p\_promo\_sk \\
	\hspace*{1em}  ${\tt\bf JOIN}$ date ${\tt\bf ON}$ cs\_sold\_date\_sk = d\_date\_sk \\
	\hspace*{1em}  ${\tt\bf JOIN}$ cd ${\tt\bf ON}$ cs\_ship\_cdemo\_sk = cd\_demo\_sk 
 }

\subsection{Aria}
\texttt{Aria} is a production service request log dataset at Microsoft that was also used in prior work~\cite{diff,storyboard}. The data set contains the following columns:\\ \texttt{records\_received\_count, records\_tried\_to\_send\_count},\\
\texttt{records\_sent\_count, olsize, ol\_w, infl, TenantId}, \\
 \texttt{AppInfo\_Version, UserInfo\_TimeZone}, \\
  \texttt{DeviceInfo\_NetworkType, PipelineInfo\_IngestionTime}. 

\begin{algorithm}[t]
\caption{Feature Selection for Clustering}
\label{alg:feat}
\begin{algorithmic}[1]
\State {\sf feats} $\gets$ (selectivity,  occurrence\_bitmap,\newline
			\hspace*{4em} $\log(x)$, $\log^2(x)$, $\min(\log(x))$, $\max(\log(x))$, \newline
             \hspace*{4em} $\overline{x}, \overline{x^2}$, std, $\min(x)$, $\max(x)$, \newline
             \hspace*{4em} \# hh, max hh, avg hh, \newline
             \hspace*{4em} \# dv,avg dv, max dv, min dv, sum dv)
\State {\sf best}  $\gets []$ \Comment{Features excluded from clustering}
\For {$i \gets 1 \to 10$}
\State {\sf feats}.shuffle() \Comment{Explore features in random order}
\State {\sf to\_exclude} $\gets  []$ 
\For {$f \in$ {\sf feats}}
\State {\sf new} $\gets$  [{\sf to\_exclude}]+[f]
\If {\textproc{ImproveCluster}({\sf to\_exclude, new})} 
\State {\sf to\_exclude} $\gets$ {\sf new}
\EndIf
\EndFor
\If {\textproc{ImproveCluster}({\sf best, to\_exclude})}
\State {\sf best} $\gets$ {\sf to\_exclude}
\EndIf
\EndFor
\State \Return {\sf best}
\end{algorithmic}
\end{algorithm}

\section{Implementation Details}
\label{app:misc}
In this section, we provide additional implementation details for the partition picker.

\subsection{Clustering}
\label{app:cluster}

\minihead{Normalization} Prior to clustering, we normalize the summary statistics to make sure that the euclidean distance is not dominated by any single statistic. We first apply a log transformation to reduce the overall skewness to all summary statistics except for selectivity estimates; for the selectivity estimates which are between 0 and 1, we use the cube root transformation instead. We then normalize each summary statistics by its average value in the training dataset. We choose the average instead of the max as the normalization factor since it is more robust to outliers. During test time, the statistics are normalized by their corresponding average values in the training dataset.

\minihead{Failure Cases} As discussed in Section~\ref{sec:cluster}, clustering does not perform well when the predicate is highly selective. Although we can use the \texttt{selectivity\_upper} feature as an upper bound for the true selectivity, in practice, we have seen that this upper bound could overestimated the true selectivity by over $10\times$ for complex predicates (see Section~\ref{sec:stats}). Therefore, we simply rely on the query semantics to estimate the complexity of the predicates. Specifically, if the predicate contains more than 10 clauses, we use random sampling instead of clustering to select sample partitions. 

\minihead{Feature Selection} We provide pseudo code for the feature selection procedure in Algorithm~\autoref{alg:feat}.

We report the features selected by the procedure on the four real-world datasets for experiments reported in \S~\ref{sec:main}: 
\begin{itemize}[topsep=0.5pt, itemsep=0.5pt,parsep=0.5pt]
\item \texttt{TPC-H*}: selectivity\_upper, selectivity\_lower, $\min(x)$, max hh, max dv, hh\_bitmap
\item \texttt{TPC-DS*}: $\log^2(x)$, $\overline{x}$, sum dv, hh\_bitmap
\item \texttt{Aria}: selectivity\_indep, selectivity\_max, $\min(\log(x))$, $\overline{x}$, $\max(x)$, avg hh, \# dv
\item \texttt{KDD}: selectivity\_indep, $\overline{x^2}$, max dv
\end{itemize}
Only a small number of features are used in each dataset, but across datasets, all four types of features are represented. This again illustrates the need for all four sketches.

Finally, we measure the quantitative impact of the feature selection procedure on clustering performance in \autoref{tab:featselect}. Similar to the experiment in \S\ref{sec:evalcluster}, we evaluate the \emph{average relative error} for estimating the query answer using different clustering procedures, and compare the total area under the error curve for different sampling budgets. Overall, feature selection consistently improves clustering performance for both clustering methods, reducing the area from 0.5\% to 15\% across datasets. 

\begin{table}[t]
\centering
\small
\caption{Area under the curve for the average relative error of clustering under different sampling budgets for Hierarchical Agglomerative Clustering (HAC) and KMeans clustering; smaller is better.}
\begin{tabular}{l|r||r|r||r|r}
\toprule
 &  HAC (ward) & +feat sel & KMeans &+feat sel\\
 \midrule
 \texttt{TPCDS}  &  4.2 & 3.8 (-9\%) &4.2 & 3.8 (-8\%) \\
\texttt{Aria}  &  2.6 & 2.3 (-14\%) & 2.7 & 2.3 (-15\%)\\
\texttt{KDD}   & .58 & .55 (-5\%) & .55 & .54 (- .5\%)\\
\bottomrule
\end{tabular}
\label{tab:featselect}
\end{table}

\subsection{Training}
We use the \texttt{XGBoost} regressor as our base model and use the squared error as the loss function. Although our models are only used for binary classification, we train them as regressors instead of classifiers. This is to address the problem that the ratio of positive to negative examples are different for different queries. Consider a query which has one partition with rows that satisfy the predicate versus a query with 100 such partitions. Missing one positive example would have a much larger impact on the final accuracy for the first query compared to the second. While a classifier can only handle class imbalance globally, with a regressor, we can scale labels differently such that the positive examples weigh more in the first query. We provide pseudo code for the training set up in Algorithm~\ref{alg:train}. 

\begin{algorithm}[t]
\caption{Training Label Generation}
\begin{algorithmic}[1]
\Require{threshold $t \in [0, 1]$, partition count $n$, feature dimension $m$, query answer dimension $d$; for each input query $i$, partition features $F_i \in \mathbb{R}^{n\times m}$ and normalized query answers on each partition $A_{i} \in [0,1]^{n\times d}$} 
\Ensure{X, Y}
\State $X \gets []$, $Y \gets []$
\ForEach {$(F_i, A_i) \in$  training} \Comment{For each query}
\State {\sf ans}  $\gets \sum(A_i)$ \Comment{Ground truth query answer}
\For{$j \gets 1 \to n$}
\State $y[j] \gets \max(A_i[j]) > t$ \Comment{Partition contribution}
\EndFor
\State {\sf  positive}  $\gets \sum y$
\For{$j \gets 1 \to n$}
\If {$y[j]$ == 1}
\State $y[j] \gets \sqrt{\frac{c}{positive}}$
\Else 
\State $y[j] \gets  -\sqrt{\frac{c}{n-positive}}$
\EndIf
\EndFor

\State $X.append(F_i)$
\State $Y.append(y)$
\EndFor
\end{algorithmic}
\label{alg:train}
\end{algorithm}

\section{Additional Results}
\label{app:result}

\subsection{Modified Learned Stratified Sampling}
\label{app:strata}
In this section, we present the three necessary modifications made to Learned Stratified Sampling~\cite{lss} in detail:
\begin{itemize}[noitemsep,topsep=0pt]
\item We move the training from online to offline, and use one trained model {\em per dataset and layout} instead of {\em per query}. LSS performs training inline for each query, using a fixed portion of the sampling budget as the training data. Training on random row-level samples may invalidate I/O gains and already require a full scan over data (\S~\ref{sec:intro}). Instead, we train the model offline on training queries sampled from the workload and use the same trained model for all test queries. 
\item We change the model's inputs and labels. LSS operates on rows, while we use partition features as inputs. LSS only considers count queries, so the label is either 0 or 1. To support aggregates and group bys, we use the partition contribution defined in \S~\ref{sec:clf} as labels.
\item We use different stratification strategies. Prior work analyzes optimal choices of strata boundaries for proportional allocation of samples, in which the sample size allocated to each stratum is proportional to its size. The analysis does not extend to our setup, so we use equi-width strata instead. To set the number of strata, we exhaustively sweep the strata sizes and select one that minimizes average relative error on the training set. We report the selected strata sizes in~\autoref{tab:strata}. 
\end{itemize} 

\begin{table}
\centering
\small
\caption{Strata sizes for the modified LSS algorithm selected via exhaustive search.}
\begin{tabular}{lrrrrrrrrr}
\toprule
& \multicolumn{9}{c}{Sampling Budget (\% data read)} \\
\cmidrule{2-10} 
 & 10 &20 & 30 & 40 & 50 & 60 & 70 & 80 & 90 \\
 \midrule
 \texttt{TPC-H*}   & 15 & 50  & 100  &  250 & 260    & 580 & 430 & 50 & 730   \\
\texttt{TPC-DS*}  & 55  &  120 &  85 &  130    &  160    & 250 & 395  & 170 &  10  \\
\texttt{Aria}  & 75     &80    & 55  & 150      &    260   & 70 & 80  & 130 & 190 \\
\texttt{KDD}   & 90   &160      & 295   & 230    &    360    & 430 & 220 & 410 & 820\\
\bottomrule
\end{tabular}
\label{tab:strata}
\end{table}

\subsection{Effect of Sampling Rate}
\label{sec:decay}

\begin{figure}[t]
\centering
\includegraphics[width=\columnwidth]{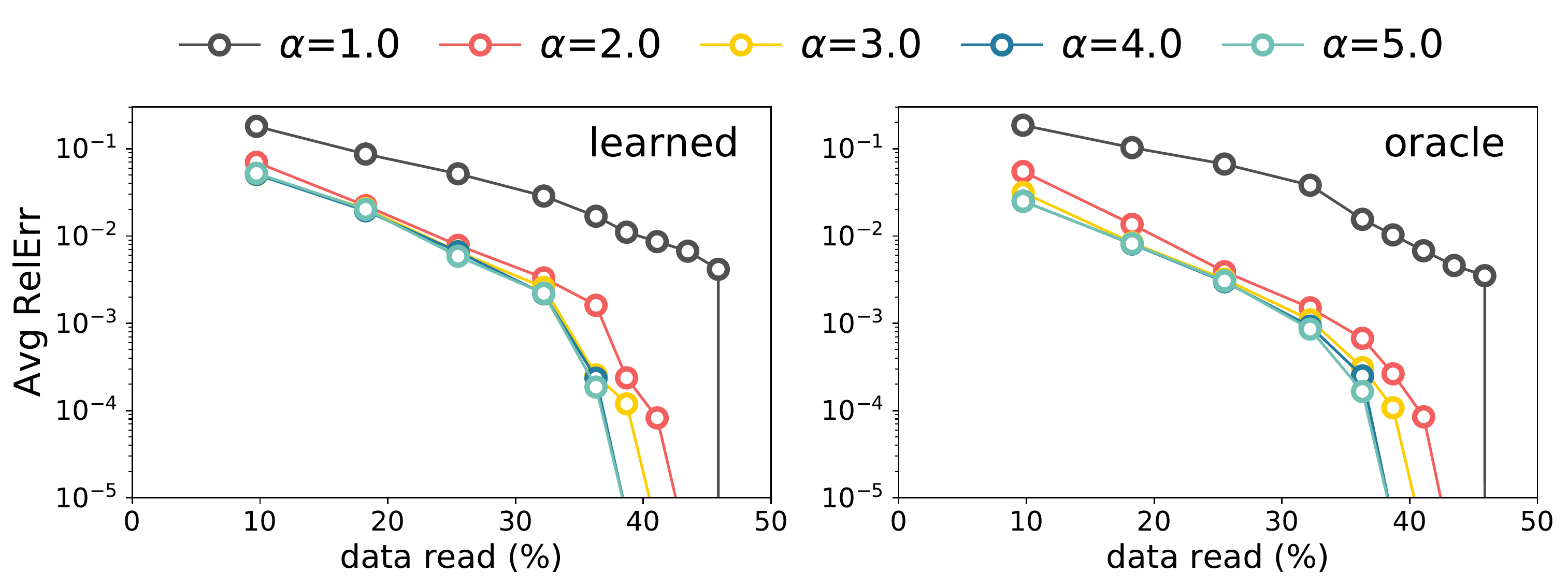}
\caption{Impact of the sampling decay rate $\alpha$ on the KDD dataset. Larger $\alpha$ improves performance, but the marginal benefits decreases. }
\label{fig:rate}
\end{figure}

We investigate the extent to which applying different sampling rates affects the performance of learned importance style sampling. Recall that we tune the sampling rate via parameter $\alpha$, which is the ratio of sampling rates between the $i^{th}$ important and the $(i+1)^{th}$ important group. The larger $\alpha$ is, the more samples we allocate to the important groups. We report the results achieved under different $\alpha$s for the \texttt{KDD} dataset (Figure~\ref{fig:rate}, left). Overall the performance improves with the increase of $\alpha$, but the marginal benefit decreases. 

We repeat the experiment and replace the trained regressors with an oracle that has perfect precision and recall (Figure~\ref{fig:rate}, right). This gives an upper bound of the improvements enabled by important-styled sampling. Compared to using learned models, the overall error decreases with the oracle, as expected. The performance gap between the learned and the oracle regressor increases with the increase of $\alpha$. The comparison shows that the more accurate the regressor, the more benefits we get from using higher sampling rates for important groups. While we used a default value of $\alpha=2$ across the experiments, it is possible to further fine-tune $\alpha$ for each dataset to improve the performance.

\subsection{TPC-H Results}
\label{app:tpch}

In this subsection, we report a detailed breakdown of the performances of \sysname and random partition-level sampling on the TPC-H queries from the generalization test (\S~\ref{sec:qgen}). To support Q8 and Q14, we rewrite the \texttt{SUM} aggregate with a \texttt{CASE} condition as an aggregate over the predicate. In addition, \sysname explicitly chooses to use random sampling instead of clustering to select samples for Q19, which has complex predicates consisting of 21 clauses (\S~\ref{app:cluster}). Our training queries are sampled randomly according to procedure described in \S~\ref{sec:qgen}. We provide an example of a randomly generated query from the \texttt{TPC-H*} schema below:\\
{\small \indent${\tt\bf SELECT}$ N1\_NAME, \\
	\indent\indent\indent\indent${\tt\bf SUM}$(L\_EXTENDEDPRICE * L\_TAX)\\
    \indent${\tt\bf FROM}$ denorm \\
    \indent${\tt\bf WHERE}$  P\_SIZE$\geq$7 ${\tt\bf AND}$ L\_COMMITDATE$\geq$``1997-09-29" \\
	\indent${\tt\bf GROUP BY }$ N1\_NAME;\\
 }

Overall, \sysname significantly outperforms random partition selection on Q1, Q6 and Q7, and performs similarly to random partition selection on other queries. In particular, Q1, Q6 and Q7 all have a small number of partitions with either rare groups or outlying aggregate values. While \sysname can identify such partitions via clustering and outlier detection, random partition selection can easily miss these important partitions especially when the sampling budget is limited.

\begin{figure*}[]
\centering
\includegraphics[width=\textwidth]{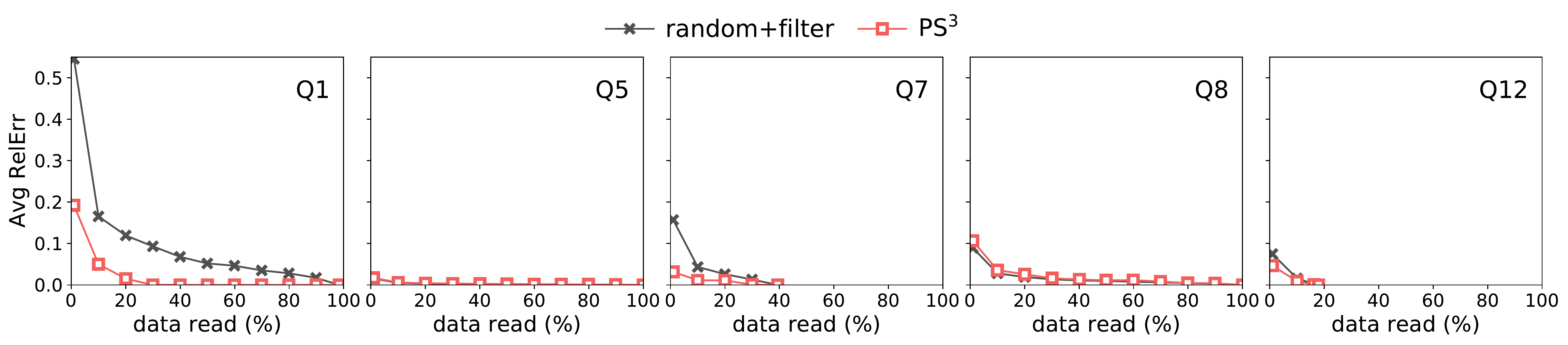}
\includegraphics[width=\textwidth]{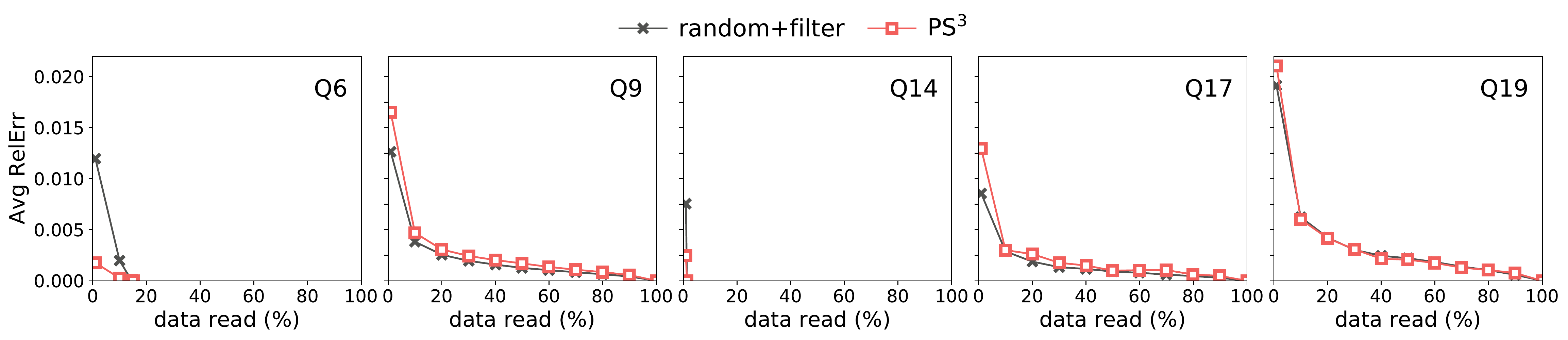}
\caption{Detailed breakdown of results on TPC-H queries used in the generalization test (\S~\ref{sec:qgen}). Overall, \sysname significantly outperforms random partition selection on Q1, Q6, Q7 and performs similarly to random partition selection on other queries. }
\label{fig:tpchall}
\end{figure*}

\section{Variance Analysis}

\subsection{Unbiased picker} 
\begin{figure}
\centering
\includegraphics[width=\columnwidth]{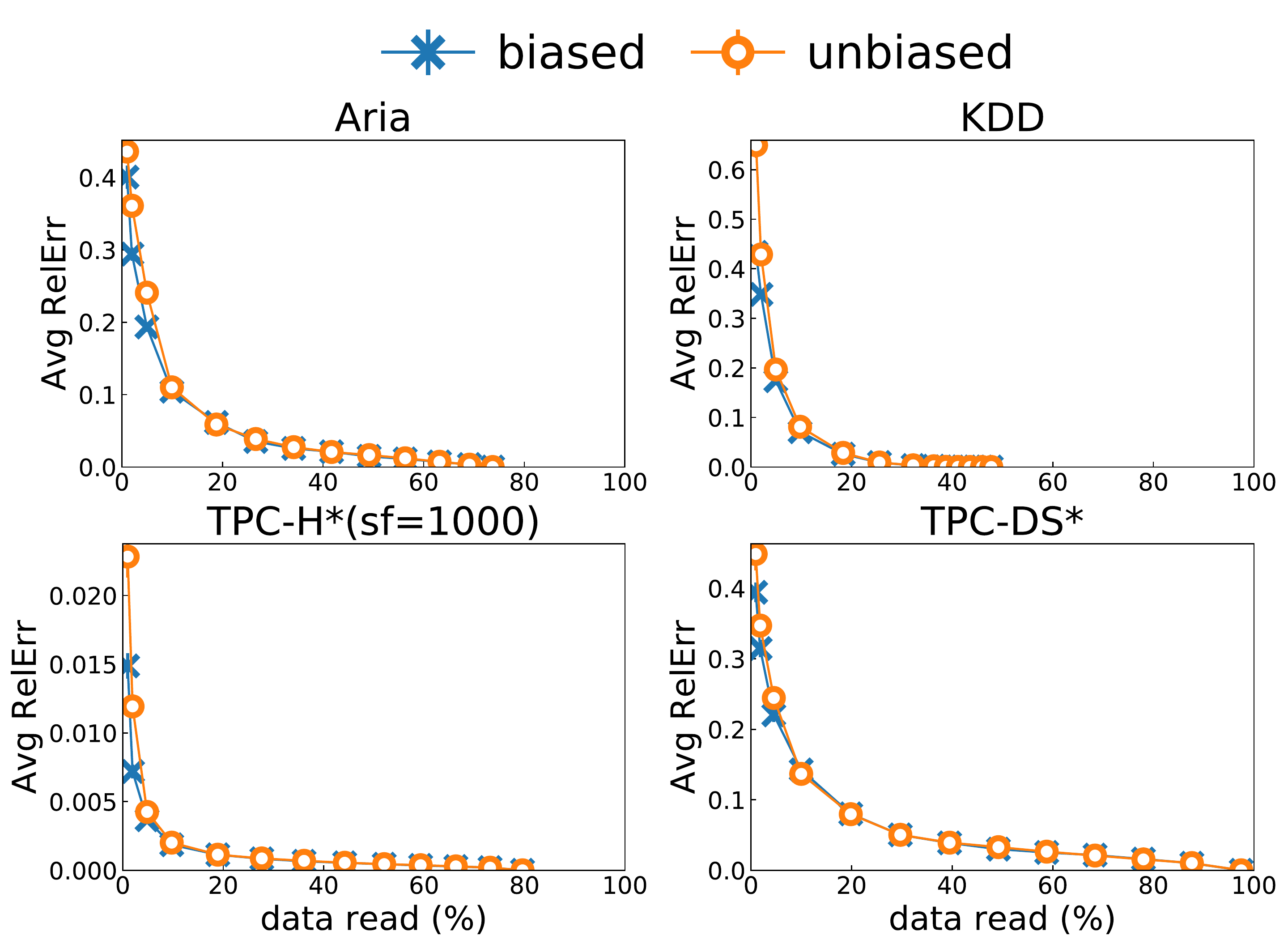}
\caption{Empirical comparison of the bias and unbiased version of the estimator. The  biased estimator tends to outperform the unbiased when the sampling fraction is small. }
\label{fig:bias}
\end{figure}

\minihead{Unbiased picker} We introduce an unbiased version of our proposed estimator that lends well to analysis. As described in \S~\ref{sec:cluster}, the biased estimator picks an exemplar partition deterministically from a cluster given the median feature vector of the cluster, whereas the unbiased estimator picks a cluster exemplar partition at random. We empirically compare the performances of the two estimators on four real-world datasets in \autoref{fig:bias}. For each test query, we run the unbiased estimator 10 times and compute the average error achieved to compare against the error achieved by the biased estimator. 

Overall, Figure~\ref{fig:bias} shows that the biased estimator achieves smaller error compared to the unbiased version when the sampling fraction is small, and that there are no significant differences in accuracy between the two estimators otherwise. In addition, for a given query, the biased version of the estimator has \emph{no variance}. Therefore, in use cases when the sampling budget is limited or when users prefer getting a deterministic answer for a given query, the biased version of the estimator might be preferred. 

\minihead{Analysis} Next, we analyze the unbiased version of the estimator using the framework of stratified sampling. Compared to a simple random sample of the same size, stratified sampling can produce an estimator with smaller variance if the elements within strata are homogeneous. In our case, each cluster is essentially a stratum; if clustering is effective, the partitions in a cluster are similar to each other, leading to a variance reduction. 

Within each stratum, we perform simple random sampling without replacement (SRSWoR) to draw a sample of size 1; the variance formula for SRSWoR can be found in Chapter 2.5.2 of~\cite{cormode2011synopses}. Note that since we only draw one sample from each cluster/stratum, in order to estimate variance of the stratum, we would need to evaluate \emph{additional} partitions per stratum. Finally, the total variance of the unbiased estimator is the sum of the variances from each stratum. 

When central limit theorem holds, the 95\% confidence interval of an estimator $Y$ is given by $\pm 1.96\sqrt{\sigma^2(Y)}$~\cite{cormode2011synopses}, where $\sigma^2(Y)$ is the variance of the estimator described above.



\subsection{Partition-level v.s. row-level sampling}
In this subsection, we compare random partition level sampling to random row level sampling. We show that under the same sampling fraction, random partition level sampling has much larger variance than random row level sampling. 

\minihead{Set up} We start with a description of the setup. For a group $G$ in the query, let $y_i$ be the value of the aggregate function on partition $i$. Let $\pi_i$ be the probability that partition $i$ is included in the sample, $\pi_{ij}$ be the probability that both partition $i$ and $j$ are in the sample, $N$ be the total number of partitions and $S$ be the set of sampled partitions. 

We wish to estimate the total value of the aggregate function for group $G$ on all partitions. For \texttt{SUM} and \texttt{COUNT} queries, the total value is $Y = \sum_{i = 1}^N y_i$. If all partitions have positive sampling probability ($\pi_i > 0, \forall i$), an unbiased \emph{Horvitz-Thompson} estimator for $Y$ under Poisson sampling is:
\[\hat{Y} = \sum_{i \in S} \frac{y_i}{\pi_i}\]
The true variance of the estimator $\hat{Y}$ is:
\begin{equation}
\sigma^2(\hat{Y}) = \sum_{i,j = 1}^{N} (\frac{\pi_{ij}}{\pi_i \pi_j} - 1) y_i y_j
\label{eqn:truevar}
\end{equation}

However, since $y_i$ is only available for partitions that are included in the sample, we can not evaluate the true variance using Eq~\ref{eqn:truevar} directly. Instead, we estimate the true variance using the sampled set of partitions $S$~\cite{cormode2011synopses}:
\begin{equation}
\hat{\sigma}^2(\hat{Y}) = \sum_{i,j = 1}^{N} (\frac{1}{\pi_i \pi_j} - \frac{1}{\pi_{ij}}) y_i y_j
\label{eqn:var}
\end{equation}

If the second-order inclusion probability $\pi_{ij} > 0$ for all pairs of partitions $i,j$, Eq~\ref{eqn:var} is an unbiased estimator for  Eq~\ref{eqn:truevar}, the true variance of $\hat{Y}$~\cite{fuller}.

\minihead{Analysis} For random partition level sampling, assume that each partition is selected in the sample with probability $p$. The expected size of $S$ is $Np$. Since the partitions are sampled independently, $\pi_{ij} = \pi_{i}\pi_{j}$. Plug the inclusion probabilities in Eq~\ref{eqn:var}, the estimator of the true variance is: 
\begin{equation}
\hat{\sigma}^2(Y_{blk}) = \sum_{i \in S} (\frac{1}{p^2} - \frac{1}{p})y_i^2
\label{eqn:block}
\end{equation}

Similarly, assume that each tuple is sampled with probability $p$. Let $t_x$ be the total value that a tuple $x$ contributes towards the aggregate for group $G$, and $S_t$ be the set of sampled tuples. Following similar derivation as Eq~\ref{eqn:block}, the estimator of the variance for random row level sampling is 
\begin{equation}
\hat{\sigma}^2(T_{row}) = \sum_{x \in S_t} (\frac{1}{p^2} - \frac{1}{p})t_x^2
\label{eqn:rowuni}
\end{equation}


Note that $y_i$ in Eq~\ref{eqn:block} is simply the sum of tuples in partition $i$. Let $b_x$ be the partition that contains tuple $x$, then $y_i =  \sum_{b_x=i} t_x$. Therefore, 
 \[y_i^2 = \sum_{b_x=i} t_x^2 + 2\sum_{\substack{x<y, \\b_x=b_y=i}}t_xt_y\] 
 Eq~\ref{eqn:block} can be rewritten as 
 \begin{equation}
\hat{\sigma}^2(T_{blk}) = \sum_{i \in S_t} (\frac{1}{p^2} - \frac{1}{p})t_i^2 + 2\sum_{\substack{i, j\in S_t,\\ i < j, b_i=b_j}} (\frac{1}{p^2} - \frac{1}{p})t_it_j
\label{eqn:row}
\end{equation}

Comparing to random row-level sampling with the same sampling fraction $p$ (Eq~\ref{eqn:rowuni}), random partition-level sampling has larger variance: Eq~\ref{eqn:row} includes an additional term that accounts for the variance contributed by tuples belonging to the same partition. 
%

\end{document}